\documentstyle[preprint,aps,epsf]{revtex}

\newcommand{\size}{14.0truecm}
\newcommand{\sizen}{7.5truecm}

\newcommand{\sizer}{9.5truecm}

\begin{document}
\preprint{BI-TP 97/55 \hspace*{1cm} JYFL 97--8}
\title{Mass number scaling in ultra-relativistic nuclear collisions
 from a hydrodynamical approach}
\author {Josef Sollfrank\footnote{present address:
Institut f\"ur theoretische Physik, Universit\"at Regensburg, Germany}}
\address{Fakult\"at f\"ur Physik, Universit\"at Bielefeld, Germany}
\author {Pasi Huovinen and P.V.~Ruuskanen}
\address{Department of Physics,
  University of Jyv\"askyl\"a, Finland \\ }
\date{\today}
\maketitle
\begin{abstract}
We study the different nucleus-nucleus collisions, O+Au, S+S, S+Ag, S+Au
and Pb+Pb, at the CERN-SPS energy in a one-fluid hydrodynamical approach
using a parametrization based on baryon stopping in terms of the
thickness of colliding nuclei. Good agreement with measured particle
spectra is achieved. We deduce the mass number scaling behaviour of the
initial energy density.  We find that
the equilibration time is nearly independent of the size of the
colliding nuclei.
\end{abstract}
\vspace*{0.2in}

\section{Introduction}\label{first}

\baselineskip 3.0ex

From heavy ion collisions at ultra-relativistic energies one would
like to learn, e.g.~how high energy and baryon number densities can
be reached in these collisions and to what degree the produced matter
is equilibrated.  Answers to these questions would tell us which
regions in the nuclear phase diagram are accessible in nuclear
collisions.  It would also show if, and to what extent the quark-gluon
phase has been reached in present SPS experiments~\cite{Satz97}.

There are no direct measurements of the achieved energy and baryon
number densities and therefore one uses models to extract the
information from the experimental data.  One commonly used approach is
hydrodynamics \cite{Strottman97} which correlates the initial densities
with measured spectra of final particles.  By definition, hydrodynamics
assumes local thermal equilibrium in the expanding matter and is
therefore unable to answer the question of thermal equilibration.

Our main goal in this work is to investigate the scaling of initial
energy density in heavy ion collisions as a function of the size of
the colliding nuclei at (nearly) fixed energy.  Therefore we study and
compare five different collisions at CERN-SPS energies, O+Au, S+S,
S+Ag, S+Au and Pb+Pb, using the same parametrization
for the initial state.  In Section
\ref{second} we introduce the parametrization of the initial
conditions.  Section \ref{third} discusses the results and in Section
\ref{fourth} conclusions are drawn.

\section{Initial Conditions}\label{second}

At ultra-relativistic energies nucleus--nucleus collisions are
expected to become transparent, meaning that the matter of final
particles is produced in a state of {\it initial collective motion}.
We have not succeeded to reproduce, even qualitatively, the baryon
rapidity distributions if we assume that the produced matter is
initially completely stopped, commonly referred to as the Landau
initial conditions.
As an example we show in Fig.~\ref{Pblandau} the
net-proton distribution in a lead-on-lead collision as obtained by
assuming that the matter of the initial fireball is at rest.
The calculation is compared with the measurement of NA49~\cite{Jacobs97}.
The disagreement between the data and the calculation is, as expected,
larger for lighter nuclei. This indicates that the
primary collision stage cannot be treated within one fluid
hydrodynamics since this would lead to compressed matter at rest.
Instead, we start the simulation after the system has equilibrated
\cite{Sollfrank97a,Ornik91}. To use this procedure we need a
parametrization of the initial conditions.  Little is known about the
initial state and a chosen parametrization can be justified only by
comparing with the data.  The parameter space of the initial
conditions, which give reasonable agreement with data turns out to be large.
Partly this results from the freedom in choosing the equation of
state.  Therefore it is difficult to pin down, e.g.~the initial
temperature achieved in a heavy ion collision.

Here we exploit the idea to fix the initial conditions using
the local thickness of colliding nuclei.  We parametrize
the nuclear stopping locally in the transverse direction in terms of
the nuclear thickness function and then constrain the energy density
from conservation of energy.  The advantage of this approach, when
combined with the geometry of the colliding nuclei, is that fixing of
the parameters fixes the initial conditions for all different
collisions at the given energy.  It also gives a better basis to
compare different collisions and to draw conclusions on the dependence
of the initial state on the nuclear size.

The main features of our parametrization of the initial conditions are
described in \cite{Sollfrank97b}. Here we summarize the most
important ingredients. Note that we consider only
impact parameter zero collisions i.e.~azimuthal symmetry is
assumed.

One of the difficulties in hydrodynamic simulations is the
treatment of the surfaces.  E.g.~in the transverse direction the outer
edges never behave hydrodynamically and even in head-on collisions a
certain fraction of nucleons never suffer a single collision.  These
nucleons are not detected in the experiment and their energy does not
contribute to the production of final matter.  To account for this we
use effective nuclear sizes, i.e.~we replace the mass numbers of
incoming nuclei, $A$ and $B$ by
\begin{equation}\label{xi}
A^{\rm eff} = \xi A$ \quad , \quad $B^{\rm eff} = \xi B
\end{equation}
and fix the geometry in terms of these effective mass numbers $A^{\rm
eff}$ and $B^{\rm eff}$.  In fixing the value of $\xi$, always close
to one, we also compensate for the few percent losses of baryon number
and energy in our numerical code.

We start out by parametrizing the rapidity distribution of baryons
$dN_{\rm B}/dy$ as a sum of projectile and target contribution for
symmetric collisions \cite{Sollfrank97b},
\begin{eqnarray}\label{param}
\frac{dN_{\rm B}}{dy}
& = &
\frac{dN^{\rm P}_{\rm B}}{dy} (x_y) + \frac{dN^{\rm T}_{\rm B}}{dy} (x_y)
= \\ && \left [C^{\rm P} \exp(ax_y^3 + bx_y^2 + cx_y) +
C^{\rm T} \exp(-ax_y^3 + bx_y^2 - cx_y) \right] (1-x_y^2) \: , \nonumber
\end{eqnarray}
using a variable $x_y = y/y_{\rm max}$, the rapidity scaled with
$y_{\rm max}= y^{\rm P}_{\rm cm}$, the projectile rapidity in the cm
frame. The factor $(1-x_y^2)$ ensures that the distribution goes
to zero at the boundary of the phase space when effects like Fermi
motion are neglected.  The functional form is motivated by the
experimental proton rapidity distributions in $p$ + $p$ collisions
\cite{Sollfrank97b} and the possibility to control the amount of
stopping. $C^{\rm P}$ and $C^{\rm T}$ are normalization constants.  The
main idea of our parametrization is that the above initial
distribution is determined locally in transverse plane.  This is
implemented by defining the parameters $a(T_A)$, $b(T_A)$ and $c(T_A)$
as functions of the local nuclear thickness
\begin{equation}\label{thickfunction}
T_A (\vec{\rho}) = \int dz \; n_{\rm B}(z, \vec{\rho}) \: ,
\end{equation}
where $n_{\rm B} (\vec{r})$ is the nuclear density for a nucleus of
mass number $A$, $z$ is the longitudinal and $\vec\rho$ the transverse
variable: $\vec r=(z,\vec\rho)$. We use the Woods-Saxon parametrization
for the nuclear density
\begin{equation}\label{ws}
n_{\rm B}(\vec{r}) = \frac{n_0}{\exp[(|\vec{r}| - R_A)/a_R] + 1} \: ,
\end{equation}
with
\begin{equation}\label{radiusdef}
R_A = 1.12 \: {\rm fm} \times A^{1/3} - 0.86
\: {\rm fm} \times A^{-1/3} \,,
\end{equation}
$a_R = 0.54$ fm and $n_0 = 0.17$ fm$^{-3}$ \cite{Bohr69}.
The functional dependence of $a$, $b$ and $c$ on $T_A$ is chosen to be
\cite{Sollfrank97b}
\begin{eqnarray}
a(T_A) & = & 1.5 \: (\sigma_{pp} \, T_{A})^{-1} \nonumber \\
b(T_A) & = & {\beta_s}\: (1 - \sigma_{pp} \, T_{A}) \nonumber \\
c(T_A) & = & 3.0  \: ,\label{abc}
\end{eqnarray}
where $\sigma_{pp} = 32$ mb is the total inelastic cross section
for $p + p$ collisions at SPS energy.
The first motivation for this parametrization is its simplicity.
Second, in the case of one collision, i.e.~$\sigma_{pp} \, T_{A} =1$,
the experimental proton rapidity distribution for $p + p$ collisions is 
recovered \cite{Sollfrank97b}.
For $\sigma_{pp} \, T_{A} > 1$, the $b$ coefficient is negative indicating
increasing stopping with growing nuclear thickness, the strength being
controlled by $\beta_s$.
It is fixed from fits to baryon rapidity spectra
in heavy ion collisions and therefore its determination includes the
hydrodynamical evolution. Changing the hydrodynamical evolution
by choosing different initial energy and (longitudinal) velocity
distributions or equation of state results in different optimal values
for $\beta_s$. In the analysis presented here the value $\beta_s = 2.25$
adequately reproduces all results.

We would like to emphasize that $T_A (\vec{\rho})$ depends on the
transverse coordinate $\vec{\rho}$ and therefore stopping
at the center is different from stopping at the edge. As an example
we plot in Fig.~\ref{inidens}
the baryon density $\gamma n_{\rm B}$ in the cm frame
for S+S collision. One sees clearly the increase in transparency with
radius. This makes the one-fluid hydrodynamical treatment of nucleus-nucleus
collisions more realistic.

For the initial local energy distribution we choose a
Gaussian form
\begin{equation}\label{enparameter}
\frac{dE(\vec\rho)}{dy} = C_{\varepsilon} \exp\left[
\frac{-(y-y_0)^2}{2\sigma_\varepsilon^2} \right] \:
\left[1 - (y/y_{\rm max})^2 \right] \: ,
\end{equation}
where the width $\sigma_\varepsilon$ and the normalization
$C_{\varepsilon}$ depend on the transverse coordinate $\vec\rho$.
The normalization $C_\varepsilon$ is given by energy conservation and
the value $y_0$ is identified with the center-of-mass
rapidity of the collision locally in the transverse plane.
Since our hydrodynamic calculation is performed in the overall
center-of-mass frame of participating nucleons, $y_0(\vec{\rho})
\equiv 0$ for zero impact parameter collisions of equal nuclei.  For
asymmetric collisions the thickness of target and projectile varies
differently with transverse radius $\rho$ and therefore the rapidity
of the local cm frame $y_0(\vec\rho)$ changes in the transverse plane.

In Eq.~(\ref{enparameter}) the only free parameter
is the width $\sigma_\varepsilon$ which
is taken to depend on the nuclear thicknesses as \cite{Sollfrank97b}
\begin{equation}\label{cy}
\sigma_\varepsilon(\rho) = \frac{c_\varepsilon}{
\left[\sigma_{pp}T_A(\rho)
\sigma_{pp}T_B(\rho)\right]^{\;\alpha_\varepsilon}}\:,
\end{equation}
where $\sigma_{pp}$ is included to make the denominator
dimensionless. The constants $c_\varepsilon$ and $\alpha_\varepsilon$
are determined from an overall
fit to the investigated collision systems with the result $c_\varepsilon =
0.8$, $\alpha_\varepsilon = 0.125$.
\footnote{These values are slightly
different from those used before in \cite{Sollfrank97b} due
to the extension to more collision systems.}

So far baryon density and energy density distributions are
specified only as functions of velocities. To convert them into spatial
distributions we have to specify the velocity field.
In the case of the Bjorken model \cite{Bjorken83}, the scaling
ansatz for the longitudinal velocity is $v_z = z/t$ and initial
conditions are usually defined at fixed proper time $\tau_0$.
Since scaling cannot hold in a finite system, we have, for numerical
convenience, chosen the rapidity $y$ instead of velocity $v_z$
to have the linear $z$--dependence:
\begin{equation}\label{vap}
y(\rho, z) = \kappa(\rho) z \,,\qquad
v_z(\rho,z) = \tanh[\kappa(\rho) z]
\:, \end{equation}
where the proportionality constant $\kappa$ depends on the
transverse radius $\vec{\rho}$ \cite{Sollfrank97b}.
For small $z$ this ansatz can approach the velocity profile
of the scaling solution. The proportionality constant is now $\kappa$
instead of $1/\tau_0$ of the Bjorken model.
We define
\begin{equation}\label{tau}
\tau_0^{\rm eff} = {1\over\kappa(\rho = 0)}
\end{equation}
as a parameter which can be regarded as an
equilibration time scale in the same way as $\tau_0$ in the scaling case.
The initial time $\tau_0^{\rm eff}$ is a parameter which is expected to
depend on the nuclear size and the collision energy and is adjusted
separately to each collision.

As described above, the model is formulated for symmetric collisions
in the center-of-mass frame of the collision.
Since we are interested in the mass number dependence of the initial state
and large part of the available data is for asymmetric collisions, we would
like to modify the parametrization (\ref{param}) to apply to
arbitrary collisions, too.

In symmetric zero impact parameter collisions the rapidity shift is the same
on the target and the projectile side and $y_0(\vec\rho)=0$ for all values of
$\vec\rho$.  For asymmetric collisions the target and the projectile
rapidities differ in the overall cm frame of the participating nucleons.
Also, since we determine the initial conditions from the nuclear thicknesses
locally in the transverse plane, we have to distinguish between the global
cm frame and the cm frame for the collision of a row of target nucleons with a
row of projectile nucleons at given $\rho$.
Eq.~(\ref{param}) is evaluated in this
local cm frame with $x_y$ defined using $y^{\rm P}_{\rm cm}$ on the projectile
side and $y^{\rm T}_{\rm cm}$ on the target side
\begin{eqnarray}\label{paramas}
\frac{dN_{\rm B}}{dy}(y)
& =  &
\frac{dN^{\rm P}_{\rm B}}{dy} (y) +
\frac{dN^{\rm T}_{\rm B}}{dy} (y) \\
& = & \left [C^{\rm P} \exp(ax_{y,{\rm P}}^3
+ bx_{y,{\rm P}}^2 + cx_{y,{\rm P}}) +
C^{\rm T} \exp(-ax_{y,{\rm T}}^3 + bx_{y,{\rm T}}^2 - cx_{y,{\rm T}}) \right]
\nonumber \\
&& \; \times \; (1+x_{y,{\rm T}})(1-x_{y,{\rm P}}) \; ,\nonumber
\end{eqnarray}
where $x_{y,{\rm T(P)}}= y/|y^{\rm T(P)}_{\rm cm}|$ on the target
(projectile) side. This ensures the important limit
that for $T_{A,B} \rightarrow \infty$ the system is stopped in the
cm frame when we use (\ref{paramas}) together with~(\ref{abc}).
Note that the distinction of local
and global cm frames is necessary also if the parametrization is used for
nonzero impact parameter collisions of equal nuclei.

We also have to modify the stopping parametrization to account for the
fact that for a fixed target size the rapidities of the nucleons of a smaller
projectile are, on the average, shifted more than those of a bigger projectile
and vice versa. This means that we must include both the target and the
projectile thickness functions into the parametrization (\ref{abc}).  Without
this modification the average momentum of projectile nucleons after the
collision with a target of fixed size would be independent of the projectile
size and the total momentum of target and projectile nucleons after the
collisions would not add up to zero in the overall momentum frame. In
principle this needs not to be the case since the mesonic degrees of freedom
can balance the momentum but we do not find it plausible that the center
of mass of nucleons would appreciable differ from the center of mass of
produced mesons.

It turns out that the momentum balance between the colliding nucleons
can be approximately satisfied if
we replace $T_A$ and $T_B$ in Eq.~(\ref{abc}) with
effective values $T_A^{\rm eff}$ and $T_B^{\rm eff}$ defined as
\begin{eqnarray}\label{asymcor}
T_A^{\rm eff} &=& T_A\: \left( \frac{T_B}{T_A} \right)^{\alpha} \nonumber \\
T_B^{\rm eff} &=& T_B\: \left( \frac{T_B}{T_A} \right)^{-\alpha} \:.
\label{taeff}
\end{eqnarray}
The factor $(T_B/T_A)^{\alpha}$ can be considered as an asymmetry correction
for the stopping. The choice of $\alpha = 0.8$ leads to very satisfactory
momentum balance within few percent for all three investigated asymmetric
systems.

We show in Fig.~\ref{inipro} the initial local baryon density
$n_{\rm B}$ and the local energy density
$\varepsilon(z,\rho)$ for the five collisions we consider. In
Fig.~\ref{inipro}a $n_{\rm B}(z,\rho)$ is plotted as
function of $z$ for $\vec{\rho} = 0$. In all cases
the stopping leads to approximately Gaussian distribution with the maximum
at the global center-of-mass position for symmetric cases and 
slightly shifted towards the heavier target side for the asymmetric cases.

Fig.~\ref{inipro}b shows the local energy density $\varepsilon(z)$
for $\vec{\rho} = 0$.
It has a Gaussian form resulting from Eq.(\ref{enparameter}) and
the linear fluid rapidity profile (\ref{vap}).
For asymmetric collisions the maximum is at the local
$y_0(\rho=0) > y_{\rm cm}^{AB}=0$. In transverse direction the
energy density is proportional to $T_A(\rho)\, T_B(\rho)$
resulting in the shapes shown in Fig.~\ref{inipro}c.

\section{Results}\label{third}

The above-described initial state is evolved in time by
solving the ideal fluid hydrodynamical equations numerically on a
2+1 dimensional grid described in detail
in \cite{Sollfrank97a} where different equations of state are compared.
Here we show results only for an equation
of state with phase transition to QGP at $T_c = 165$ MeV labeled
as EOS A in \cite{Sollfrank97a}.

Freeze-out occurs when the mean free path of particles is of the
same order than  the size of the fireball. For the collisions of light nuclei 
this criterium leads to freeze-out temperatures of
$T_f \approx 140$ MeV \cite{Goity89}. Instead of taking contours
of constant temperature we define the freeze-out on a space-time 
surface of constant energy density of $\varepsilon_{f} = 0.15 \; 
{\rm GeV/fm}^3$. This results in an average freeze-out 
temperature of $T_f \approx 140$ MeV.
The freeze-out contours 
are plotted in Fig.~\ref{freeze} for
the different collisions we have studied. An increase in the size of the 
system leads to longer lifetimes as expected. One sees further that in
asymmetric collisions the heavier target side 
lives longer.

For Pb+Pb collisions we find indications (see below) that the
canonical freeze-out temperature, $T_f \approx$140 MeV, might be too high 
for this larger system. Therefore 
we calculated the lead-on-lead collision also with freeze-out
at $\varepsilon_{f} = 0.069 \; {\rm GeV/fm}^3$ corresponding
to an average freeze-out temperature of $T_f \approx 120$ MeV.

\subsection{Spectra}

Particle spectra are calculated using the prescription of
Cooper and Frye \cite{Cooper74} including the same hadrons
as in the construction of the equation of state \cite{Sollfrank97a}.
In the discussion of particle spectra we concentrate on the most
abundant particles, i.e.\ negative particles, pions
and (net) protons because for them the hydrodynamical treatment
with assumed chemical equilibrium is most reliable.
We show in Fig.~\ref{negrap} the
rapidity distributions of negative particles for the various collisions.
A good agreement is seen in these rapidity spectra.
To a small extent the agreement in the normalization is improved
by adjusting the $\xi$ parameter in (\ref{xi}) which accounts for the
impact averaging. This parameter is determined for each collision
system separately by optimizing the fit to the negative particles, but as
shown in Table \ref{tab1} the variations of $\xi$ are quite small.
The calculation for Pb+Pb collisions with the lower freeze-out
temperature of 120 MeV is shown as dotted line in 
Figs.~\ref{negrap}--\ref{npropt}.

The rapidity spectra of negative particles are most sensitive to the
initial energy distribution.  Our parametrization (\ref{enparameter})
which includes only two free parameters $c_\varepsilon$ and
$\alpha_\varepsilon$ in the expression of the width (\ref{cy})
reproduces all five measured rapidity spectra of negative particles.  This
suggests that (\ref{enparameter}) could be used for any
collision system at SPS energies.  One should keep in mind, however,
that the optimal values for the parameters $c_\varepsilon$ and
$\alpha_\varepsilon$ in Eq.~(\ref{cy}) depend on the equation of state
used for the hydrodynamical evolution.

The rapidity spectra of net protons are given in Fig.~\ref{nprorap}.
For S+S the $p - \bar{p}$ distribution shows two maxima in the target
and projectile fragmentation region which are nicely reproduced by the
calculation. For Pb+Pb collisions stronger stopping is expected.
The choice of parameters in Eq.~(\ref{abc}) together with the
EOS is able to account for the larger stopping in a
quantitatively satisfactory way as
seen from the comparison with preliminary data from NA49
\cite{Jacobs97}.

In the asymmetric collisions the net proton rapidity distributions are
reproduced only around midrapidity and in the forward region while
at target fragmentation region the net proton yields are
underestimated considerably.  The reason for this deviation
can be understood by considering the geometry of an asymmetric
collision. The initial state of the hydrodynamical calculation is
determined by using the cylindrical volume as
cut out of the target by the smaller projectile nucleus.
However, during the first collisions the
projectile nucleons moving through the target
acquire transverse momentum, some of them penetrate into the region
outside the cylinder and interact further.
Similarly some of the produced particles can interact with the nucleons
outside the cylinder. Nucleons from these interactions are not included in the
hydrodynamic description but are observed experimentally in the target
fragmentation region. It is reasonable to assume that
these additional protons will not play any role in the hydrodynamical
calculation. This assumption is supported by the $\Lambda$ rapidity
distributions in Fig.~\ref{nprorap} which show good agreement between
the calculated and measured spectra even in the target fragmentation region.

In Fig.~\ref{negpt} we show that the measured transverse momentum
distributions of negatives and $\pi^0$'s agree quite well with the
calculated spectra except in the case of O+Au collisions, where the
large low-$p_T$ enhancement cannot be reproduced
and the calculation exhibits slightly too much flow.
For Pb+Pb the very forward pions are flatter than in the calculation.
The particle density at this edge of the phase space, however, is
becoming so small that hydrodynamics with relatively strong longitudinal
flow can lead to an artificially large transverse cooling.
Thus we should not expect that hydrodynamics can describe the far edges
of fragmentation regions well.

A closer look on the experimental transverse spectra shows that nearly
all of them have a small but visible concave shape in $m_T - m_0$
which is not reproduced by the calculation. The calculation
gives only the average slope right. We think that the explanation
is the too simple freeze-out procedure applied in the calculation.
In fact, in such a small system as created in the nuclear collisions,
the last interaction of a final hadron can take
place with non-zero probability
at any space-time point. Therefore the spectra receive contributions
from all temperatures and not only from the assumed freeze-out
temperature. Note that the number of particles in the tail of transverse
distribution is very small indicating that rare earlier escapes of
high--$p_T$ particles would not invalidate the hydrodynamic treatment
of the bulk of the final matter. Improvements have been
suggested \cite{Grassi95} but they are difficult to
implement technically.

Finally, we show in Fig.~\ref{npropt} the transverse momentum
distribution of heavy particles.  In S+S collisions the net proton
distribution is in agreement with the data supporting the assumption
of freeze-out at a common transverse flow velocity with pions.  For
S+Ag and S+Au we obtain agreement in the $p_T$-spectra of $\Lambda$'s.
The agreement in the normalization should be considered accidental.
It is obtained with a freeze-out temperature of $T_f\approx 140$ MeV
and assuming full chemical equilibration.  More detailed studies on
chemical equilibration indicate higher chemical freeze-out
temperatures but not full chemical equilibration of strangeness
\cite{Sollfrank97c,Becattini97}.  In our calculation the normalization
is achieved because the lower freeze-out temperature is compensated by
the assumption of full chemical equilibration of $\Lambda$'s. In the
transverse spectrum the lower freeze-out temperature is balanced by
the transverse flow.

For Pb+Pb collisions the calculated transverse spectrum of net protons 
is slightly steeper than the data even at central rapidities  
if a freeze-out temperature $T_f \approx 140$ MeV
is used. This was already observed in \cite{Schlei97} and is an
indication of the freeze-out taking place at lower temperature 
in the larger Pb+Pb system.
Therefore we redid the calculation with a freeze-out temperature 
$T_f\approx 120$ MeV as suggested in \cite{Kampfer96,Hung97,Wiedemann97}.
The result is shown as dotted line in
Fig.~\ref{npropt}. The agreement for the transverse spectrum of net protons
is clearly improved. The corresponding $m_T$-spectrum of negatives in
Fig.~\ref{negpt} is only slightly modified compared to the freeze-out at
$T\approx 140$ MeV. 

We did a calculation for S+S collisions with $T_f = 120$ MeV, too.
While the $p_T$ slope of negatives still agrees with the data, that
of net-protons, and in particular that of kaons and Lambdas, 
comes out flatter than the data.

We conclude that our calculations support the kinetic estimates for the 
freeze-out \cite{Goity89}, especially its dependence on the size of 
the system. However, the freeze-out temperatures favoured by the 
$p_T$-spectra are model dependent since we found that results for 
Pb+Pb collisions with $T_c=200$ MeV and $T_f\approx 140$ MeV
are close to those with $T_c=165$ MeV and $T_f\approx 120$ MeV. 

\subsection{Parameter Systematics}

We next discuss the $A$-dependence of various quantities
in our model. In Table \ref{tab1} we show parameters which define the
collision (upper part) and  describe the initial and freeze-out state.
The parameters $\xi$ and $\tau^{\rm eff}_0$ have been adjusted
to each collision system. $\xi$ is always close to one
showing that the
experiments were triggered to most central collisions and include
nearly the whole energy and baryon number. The symmetric collisions
show a trend to a lower value of $\xi$ indicating that the impact
averaging is more important for these cases.

The $\tau^{\rm eff}_0$ determines the slope of the initial rapidity
profile (\ref{tau}) as function of $z$ and thus the spatial
distributions of densities.  A simultaneous fit to the longitudinal
and transverse spectra determines the optimal $\tau^{\rm eff}_0$.  It
turns out that $\tau^{\rm eff}_0$ is very similar,
$\tau^{\rm eff}_0\sim 1.3$~fm$/$c, for all collisions
from S+S to Pb+Pb. However, the sensitivity of final spectra on
$\tau^{\rm eff}_0$ is not very strong,
since the density parametrizations are done in rapidity space.  Within
a variation of $\pm 0.2$ fm/c in $\tau^{\rm eff}_0$ the fits remain
acceptable.

As mentioned already earlier, $\tau^{\rm eff}_0$ can be considered to
give an estimate of the thermalization time.  On the other hand the
initial longitudinal extension can be related to the equilibration
time for the whole system in the global center-of-mass frame. The
longitudinal extension of all systems is roughly the same and leads to
an estimate of a common initial time of around 3 fm/c as counted from
the time when the front edges of the nuclei meet to the time when all
quanta from primary collisions have formed and reached approximate
equilibrium.  The difference compared to $\tau^{\rm eff}_0$ arises
mainly from the finite thickness of colliding nuclei. The numbers are
consistent with the findings in the three-fluid hydrodynamical
studies.  They indicate a pressure equilibration time to be of the
order of 1.5 fm/c for Pb+Pb collisions \cite{Brachmann97} at SPS
energies.  In the hydrodynamic treatment, carried out in the
center-of-mass system of participating nuclei, the initial conditions
are specified at global equilibration time of the whole system which,
however, does not enter the calculation as an explicit parameter.

In Table \ref{tab1} we give values of the energy density
$\varepsilon$ of the initial fireball formed in different collisions.
We average over the transverse area in a unit rapidity interval at $y=0$,
\begin{equation}\label{avr}
\overline{\varepsilon}_{R} = \frac{\int d\rho \: \rho \:
\varepsilon(\rho)}{\int d\rho \: \rho}\,,
\end{equation}
in order to compare with various estimates, in particular with those
from the experimental transverse energy flow.

We expect the $A$-dependence of the initial energy density to lie
between the following two crude estimates based on the assumption
that $\varepsilon$ is proportional to the number of sources in a
nucleus-nucleus collision. An upper limit is provided by
the possibility that each nucleon in the
projectile can interact with any nucleon on its path through the
target and v.v.\ with probabilities given, e.g.\ by the  Glauber model
\cite{Glauber59}. The assumption that the energy of the nucleons is
degraded
independently in each collision with a certain fraction transferred
to the fireball, leads to the scaling of the energy density as
 \begin{equation}\label{glauber}
\varepsilon^{\rm Gl}
\propto T_A \: T_B \propto R_A \:R_B \:
\propto (AB)^{1/3}.
\end{equation}
The other extreme is the wounded nucleon model \cite{Bialas76}
which assumes that each nucleon contributes as a single source, see
also \cite{Bjorken83}. Then the number of sources on the symmetry
axis is proportional to $T_A +T_B $ with a
scaling behaviour, which for the mass numbers of interest, can be
approximated with the form
\begin{eqnarray}
\varepsilon^{\rm WN}
&\propto& T_A  + T_B \: \propto R_A + R_B \nonumber \\
&\propto& A^{1/3} + B^{1/3} = (AB)^{1/6} \left[ (A/B)^{1/6} + (B/A)^{1/6}
\right] \approx 2(AB)^{1/6} \: .
\label{bjorken}
\end{eqnarray}
It should be kept in mind that the physics of particle production is
expected
to depend on the collision energy. Soft processes can dominate
at the SPS energy but at RHIC and in particular at LHC the hard
and semi-hard production of jets and minijets will probably be the
main mechanism \cite{Eskola94,XNWang97}.
This means that the mass number scaling of $\varepsilon$ will be
different at different energies.

We investigate the scaling behaviour of the initial energy density
by fitting the $A$-dependence with
\begin{equation}\label{efit}
\varepsilon = \varepsilon_0 \; (A^{\rm eff}B^{\rm eff})^{\gamma} \: .
\end{equation}
The fit to the transversely averaged calculated values
$\overline{\varepsilon}_{R}$ (open circles) against the product
of effective mass numbers
$A^{\rm eff}B^{\rm eff}$ (cf.\ Eq.~(\ref{xi})) is shown in a double
logarithmic plot in Fig.~\ref{sl}. The minimum $\chi^2$-fit to the
five collision systems results in $\gamma = 0.22$.

The result is between the two examples mentioned above (cf.\
Eq.~(\ref{glauber}) and (\ref{bjorken})), somewhat closer to the
wounded nucleon alternative than the Glauber type behaviour.  The
classical Glauber treatment (\ref{glauber}) neglects possible
interference effects like the Landau--Pomeranchuck--Midgal effect
\cite{Landau53} which could lead to a weaker dependence on the mass
numbers.  On the other hand, in the wounded nucleon picture the
effects of multiple interactions may require an increase of the source
strength per nucleon leading to faster growth than the approximate
$(AB)^{1/6}$ behaviour.

Our initial energy density may also be compared with the estimates
from the experimental transverse energy flow
(or multiplicity densities) using the
Bjorken model for the initial volume \cite{Bjorken83}:
\begin{equation}\label{ebjorken}
\varepsilon_{\rm Bj} = \frac{({\rm d} E_T/{\rm d}\eta)_{\rm max}}{
\pi R^2_{\rm proj} \tau_0}\,.
\end{equation}
These results for $\tau_0 = 1$ fm/c, taken from \cite{Bachler91} and
\cite{Margetis95}, are also shown in Fig.~\ref{sl}
together with the fit, Eq.~(\ref{efit}).
The $\chi^2$-fits for the $\overline{\varepsilon}_{R}$ values in our
model and the Bjorken estimates from experiment are:
\begin{eqnarray}
\overline{\varepsilon}_{R} &=& 0.63 \: {\rm GeV/fm}^3 \:
\left(A^{\rm eff}B^{\rm eff}\right)^{0.22} \\
\varepsilon_{\rm Bj} &=& 0.55 \: {\rm GeV/fm}^3 \:
\left(A^{\rm eff}B^{\rm eff}\right)^{0.16} \: .
\end{eqnarray}
It is interesting to notice that $\varepsilon_{\rm Bj}$ has the
scaling of $(AB)^{1/6}$ as expected from the wounded nucleon picture.

We see in Fig.~\ref{sl}
that our average initial energy density is roughly a factor of
two higher than the Bjorken estimate. This difference reflects the
assumed hydrodynamic behaviour. In the hydrodynamic expansion the
thermal energy is transferred to collective flow motion by the
expansion work of pressure.  When the flow is asymmetric in
longitudinal and transverse directions the stronger longitudinal flow
leads to larger energy transfer into the longitudinal degrees of
freedom. As a result the energy per unit rapidity is decreased in the
central rapidity region and increased in the fragmentation regions.
If we calculate the initial energy density using the Bjorken estimate,
Eq.~(\ref{ebjorken}) from our calculated final spectra, we obtain
numbers which are consistent with the estimates from experiment.

We would like to point out that the scaling of energy density is not
very sensitive to the parameters used in Eq.~(\ref{param}) for the
initial baryon distributions.  The initial energy density is
dominantly given by the $\tau^{\rm eff}_0$ and the parametrization of
the width $\sigma_\varepsilon$ (\ref{cy}) in the Gaussian energy
distribution (\ref{enparameter}).  This parametrization is fitted to
the rapidity spectra of negative particles.  Therefore the scaling in
the initial energy density is mainly extracted from the experimental
pion spectra.

We next discuss quantities which characterize the freeze-out. The lifetime
$t_f$ of the fireball, defined as the freeze-out time at the center,
increases as expected with the size of the fireball.  We see a doubling of
the lifetime when going from S+S to Pb+Pb indicating approximately linear
dependence on the nuclear radius.

An important quantity from the point of view of the hadron spectra is the
transverse flow velocity at freeze-out. In a local hydrodynamical
calculation the transverse flow is different for different fluid
cells. In order to compare with average flow velocities of
simpler models \cite{Schnedermann93} or phenomenologically extracted values
\cite{Bearden96} we have to average over the relevant range of the
freeze-out surface. At central rapidities
we choose for convenience to average over all fluid cells with
rapidity $|y| < 0.25$ in the global fireball rest frame.
This corresponds to the region of highest transverse flow.
The resulting values for $\langle v_\rho \rangle$ are given in
Table \ref{tab1}. Note that the range $|y| < 0.25$ does not
necessarily correspond to the rapidity region where the transverse
momentum spectra in Figs.~\ref{negpt} and \ref{npropt} are measured.

The numbers for $\langle v_\rho \rangle$ are very similar with the
exception of S+S. The fact that we get for O+Au a rather similar
average flow velocity than for Pb+Pb
is a result of taking the same freeze-out energy density independent
of collision size. Using for Pb+Pb collisions the lower
freeze-out energy density corresponding to $T_f\approx 120$ MeV
we find a considerable stronger flow than in the
smaller systems. A system-size dependent freeze-out can naturally
be incorporated if one uses a dynamical freeze-out criteria
as suggested in \cite{Schnedermann93,Hung97} resulting in a
decrease of freeze-out temperature with increasing nuclear size.

The flow values for the symmetric collisions may be compared with the data
analysis of NA44 \cite{Bearden96}.  They extracted the mean transverse
velocity from a fit to various hadron spectra obtaining $\langle
v_\rho \rangle = 0.24 \pm 0.10 $ for S+S and $\langle v_\rho \rangle
= 0.36 \pm 0.14$ for Pb+Pb \cite{Bearden96} in agreement with our
studies. Our freeze-out temperature $T_f\approx 140$ MeV agrees with their
fitted value $T_f = 142 \pm 5$ MeV for S+S. But for Pb+Pb they
favour a somewhat higher value $T_f = 167 \pm 13$ MeV while our results
favour the lower freeze-out temperature of 120 MeV
leading to $\langle v_\rho \rangle = 0.44$ for Pb+Pb
in agreement with \cite{Kampfer96}.

\section{Conclusions}\label{fourth}

We have constructed a model for initial conditions in nucleus-nucleus
collisions at CERN-SPS energies based on the parametrization of baryon
stopping by specifying the initial baryon number distribution
$dN_{\rm B}/dy$. The parametrization is expressed as a function of
the local thickness of colliding nuclei in the transverse plane.
The major advantages of this
approach are that it reduces the arbitrariness of choosing the initial
conditions for different collisions and allows a proper treatment of
transverse geometry of the colliding nuclei.
We think that this is an important improvement compared to the earlier
approaches  \cite{Sollfrank97a,Ornik91} where different collisions are
not related and the longitudinal dependence
is treated equally both on the symmetry axis at $\rho=0$ and at the outer
edge ($\rho = R_{\rm P}$) of the collision system.

The success in reproducing the single particle spectra for various
collisions gives confidence that the deduced mass number dependencies
of the initial and final state quantities are reasonable.  An
important example is the scaling of the the initial energy density.
Even though the observed dependence of hadron multiplicity on the mass
number is close to linear, corresponding to the $(AB)^{(1/6)}$ scaling
for the density, the initial energy density grows faster as shown in
Fig.~\ref{sl} reflecting the effect of hydrodynamic expansion on the
final transverse energy flow.  For the S+Au and Pb+Pb collisions
we obtain a ratio $\varepsilon_{S+Au}/\varepsilon_{Pb+Pb}\approx 2/3$
which, in addition to the volume effects, could be related to the
observed features of J/$\Psi$ suppression \cite{Gonin96}.  In our
parametrization the center of the fireball starts out in the
quark-gluon plasma phase for all considered collisions.  However, we
consider the scaling more reliable than the absolute value of energy
density which is more sensitive on the details of the parametrization.
One should also note that in smaller systems the high density occurs
only for a short time interval in a small fraction of the total
volume.

For further improvements in determining the initial conditions of
nuclear collisions a better understanding of the freeze-out physics is
needed, e.g.\ to allow emission from the whole space--time
volume~\cite{Grassi95}.  For a proper treatment also a sequential
freeze-out for various particle species should be implemented
\cite{Hung97}.  Since the particle ratios are a result of the whole
evolution and not only of the freeze-out conditions this would help more
directly in pinning down the initial state in nuclear collisions.

\section*{ACKNOWLEDGMENT}
This work was supported by the Bundesministerium f\"ur Bildung und
Forschung (BMBF) under grand no.\ 06~BI~556~(6) and 06~BI~804 and by
the Academy of Finland grant 27574. We gratefully acknowledge helpful
discussions with M. Prakash, R. Venugopalan, M. Ga\'zdzicki, U. Heinz,
J. Rafelski and H. Sorge.


\newcommand{\IJMPA}[3]{{ Int.~J.~Mod.~Phys.} {\bf A#1}, #3 (#2)}
\newcommand{\JPG}[3]{{ J.~Phys. G} {\bf {#1}}, #3 (#2)}
\newcommand{\AP}[3]{{ Ann.~Phys. (NY)} {\bf {#1}}, #3 (#2)}
\newcommand{\NPA}[3]{{ Nucl.~Phys.} {\bf A{#1}}, #3 (#2)}
\newcommand{\NPB}[3]{{ Nucl.~Phys.} {\bf B{#1}}, #3 (#2)}
\newcommand{\PLB}[3]{{ Phys.~Lett.} {\bf {#1}B}, #3 (#2)}
\newcommand{\PRv}[3]{{ Phys.~Rev.} {\bf {#1}}, #3 (#2)}
\newcommand{\PRC}[3]{{ Phys.~Rev. C} {\bf {#1}}, #3 (#2)}
\newcommand{\PRD}[3]{{ Phys.~Rev. D} {\bf {#1}}, #3 (#2)}
\newcommand{\PRL}[3]{{ Phys.~Rev.~Lett.} {\bf {#1}}, #3 (#2)}
\newcommand{\PR}[3]{{ Phys.~Rep.} {\bf {#1}}, #3 (#2)}
\newcommand{\ZPC}[3]{{ Z.~Phys. C} {\bf {#1}}, #3 (#2)}
\newcommand{\ZPA}[3]{{ Z.~Phys. A} {\bf {#1}}, #3 (#2)}
\newcommand{\JCP}[3]{{ J.~Comp.~Phys.} {\bf {#1}}, #3 (#2)}
\newcommand{\HIP}[3]{{ Heavy Ion Physics} {\bf {#1}}, #3 (#2)}

{}

\newpage

\noindent {TABLE \ref{tab1}.~~}
Summary of parameters characterizing different nuclear collisions.
A QGP equation of state with $T_c=165$ MeV
(EOS A \cite{Sollfrank97a}) is used.
Quantities in the first part characterize the colliding system, those in
the second part the initial conditions of produced matter
where values of $\xi$ and $\tau^{\rm eff}_0$ are obtained from a fit
to each specific collision, and the quantities in the third part describe
the system at the freeze-out. For Pb+Pb collisions the numbers in
parenthesis correspond to a freeze-out energy density
$\varepsilon_{f} = 0.069 \; {\rm GeV/fm}^3$ with an average freeze-out
temperature of 120 MeV. The Bjorken
estimates $\varepsilon_{\rm Bj}$ are deduced from experimental
${\rm d}E_T/{\rm d} \eta$ distributions taken from
\cite{Bachler91} and \cite{Margetis95}.

\begin{center}
\begin{tabular}{|l|c|c|c|c|c|}
\hline
collision & S + S     &  O + Au  & S + Ag  & S + Au  &  Pb + Pb \\
$B + A$    & 32 + 32  & 16 + 197 & 32 + 108& 32 + 197& 207 + 207 \\
lab. energy (GeV/A)&200& 200     & 200     & 200     & 158 \\
$y_{\rm cms}$  &  3.03       &  2.54      &  2.73    &  2.62   &  2.92   \\
\hline
$\xi      $  & 0.9       &   1.0   & 0.9     &  0.95   &   0.9  \\
$\tau^{\rm eff}_0$ (fm/c)&1.2& 1.4 & 1.4     &  1.5   &   1.3  \\
$\overline{\varepsilon}_{R} $ (GeV/fm$^3$)
             &  2.5      &   4.1     & 3.8     &  4.8   &  6.7   \\
{\rm exp.} $\varepsilon_{\rm Bj}$ {\rm (GeV/fm$^3$)}
             &  1.3 & 2.3 & 2.1 & 2.6 & 3.0   \\
$\overline{T}_{R} (z=0)$ (MeV)
             &  187       &  203      &  202    &  212   &  227   \\
$R(\xi A)$ (fm)
             &  3.15       &  6.37    & 4.96     &  6.37   &  6.25   \\
$R(\xi B)$ (fm)
             &  3.15       &  2.48    & 3.15     &  3.28   &  6.25   \\
\hline
cent. $t_f$ (fm/c)
             &  6.3       &  6.5    & 7.5     &  8.0   &  11.8 (14.7) \\
$\langle v_\rho \rangle (|y_z|<0.25)$ (c)
             &  0.27       &  0.34    & 0.31     &  0.35   &  0.33  (0.44)\\
\hline
\end{tabular}
\end{center}
\newpage


\newpage

\section*{Figure Captions}

\noindent {Fig. \ref{Pblandau}.~~}
Rapidity spectra of $p-\bar p$ in Pb+Pb collisions for the Landau
initial conditions with no initial flow velocity.  The calculated
spectrum is compared with preliminary data from NA49 \cite{Jacobs97}.

\vspace*{0.2in}

\noindent {Fig. \ref{inidens}.~~}
Initial baryon density distribution $\gamma n_{\rm B}$ in the
$z$-$\rho$ plane for a S+S collision plotted in the overall
center-of-mass frame at fixed initial time.

\vspace*{0.2in}

\noindent {Fig. \ref{inipro}.~~}
Initial density distributions for
S+S (solid), O+Au (dotted), S+Ag (dot-dashed), S+Au (dashed) and
Pb+Pb (solid) collisions. In panel (a)
the baryon density  $n_{\rm B}$ and in (b) the
energy density  $\varepsilon$ are shown as functions of $z$ on the
symmetry axis $\vec{\rho}=0$. In panel (c) the radial dependence of
energy density $\varepsilon$ is shown at $z=0$.

\vspace*{0.2in}

\noindent {Fig. \ref{freeze}.~~}
Freeze-out contours $\varepsilon_f = 0.15$ GeV/fm$^3$ for the
various collisions. The lines are as in Fig.~\ref{inipro}.

\vspace*{0.2in}

\noindent {Fig. \ref{negrap}.~~}
Comparison of rapidity spectra of negative particles
measured in various collisions with our results for
EOS~A \cite{Sollfrank97a}. The average freeze-out temperature is
140 MeV (solid line) and 120 MeV (dotted line). The data sets are:
S+S (NA35) \cite{Bachler94}; 
Pb+Pb (NA49) \cite{Jones96};
O+Au (NA35) \cite{Alber97};
S+Ag (NA35) \cite{Gazdzicki95} 
and S+Au (NA35) \cite{Gazdzicki95}.
The Pb+Pb data is preliminary.

\vspace*{0.2in}

\noindent {Fig. \ref{nprorap}.~~}
Comparison of rapidity spectra of $p-\bar p$ and $\Lambda$
measured in various collisions with our results for
EOS~A \cite{Sollfrank97a}. The average freeze-out temperature is
140 MeV (solid line) and 120 MeV (dotted line).
The data sets are:
$p-\bar{p}$ in S+S (NA35) \cite{Bachler94}; 
$p-\bar{p}$ in Pb+Pb (NA49) \cite{Jacobs97}; 
$p-\bar{p}$ in S+Ag (NA35) \cite{Rohrich94}; 
$p-\bar{p}$ in S+Au (NA35: $y > 3$) \cite{Rohrich94}
(two data points with $y < 3$ are proton data from NA44 \cite{Murray95});
$\Lambda$ in S+Ag (NA35) \cite{Alber94} and 
$\Lambda$ in S+Au (NA35) \cite{Alber94}.
The Pb+Pb data is preliminary.

\vspace*{0.2in}

\noindent {Fig. \ref{negpt}.~~}
Comparison of transverse momentum spectra of negative particles
and $\pi^0$ measured in various collisions with our results for
EOS~A \cite{Sollfrank97a}. The average freeze-out temperature is
140 MeV (solid line) and 120 MeV (dotted line).
The data sets are:
S+S (NA35) \cite{Bachler94}; 
Pb+Pb (NA49) \cite{Jones96} for rapidity intervals of width 0.5
  and center at (top to bottom) 3.4, 3.9, 4.4, 4.9, 5.4 and
  data sets successively scaled down by $10^{-n}$;
O+Au (NA35) \cite{Alber97};
S+Ag (NA35) \cite{Rohrich94} for rapidity intervals (top to bottom)
  $0.8 \le y \le 2.0$, $2.0 \le y \le 3.0$, $3.0 \le y \le 4.0$,
  and $4.0 \le y \le 4.4$ and data sets successively scaled
  by $10^{-n}$ ($n = 0$,1,2,3); 
S+Au (WA80) \cite{Santo94}.
The Pb+Pb data is preliminary.

\vspace*{0.2in}

\noindent {Fig. \ref{npropt}.~~}
Comparison of transverse momentum spectra of $p - \bar{p}$ and
$\Lambda$ measured in various collisions with our results for
EOS~A \cite{Sollfrank97a}. The average freeze-out temperature is
140 MeV (solid line) and 120 MeV (dotted line).
The data sets are:
$p - \bar{p}$ in S+S (NA35) \cite{Bachler94}; 
$p - \bar{p}$ in Pb+Pb (NA49) \cite{Jones96} 
  for rapidity intervals of width 0.5
  and center at (top to bottom) 2.9, 3.4, 3.9, 4.4, 4.9, 5.4 and
  data sets successively scaled down by $10^{-n}$;
$\Lambda$ in S+Ag (NA35) \cite{Alber94} and 
$\Lambda$ in S+Au (NA35) \cite{Alber94}. 
The Pb+Pb data is preliminary.

\vspace*{0.2in}

\noindent {Fig. \ref{sl}.~~}
Initial average energy density plotted against
$A^{\rm eff}B^{\rm eff}$. Open circles show the calculated values
Eq.~(\ref{avr}) and the diamonds the
Bjorken estimates Eq.~(\ref{ebjorken}) from experimental data
\cite{Bachler91,Margetis95} with lines
corresponding to minimum $\chi^2$ fits of the form Eq.~(\ref{efit}).

 \newpage
\begin{center}
   \begin{minipage}[t]{\size}
         \epsfxsize \size \epsfbox{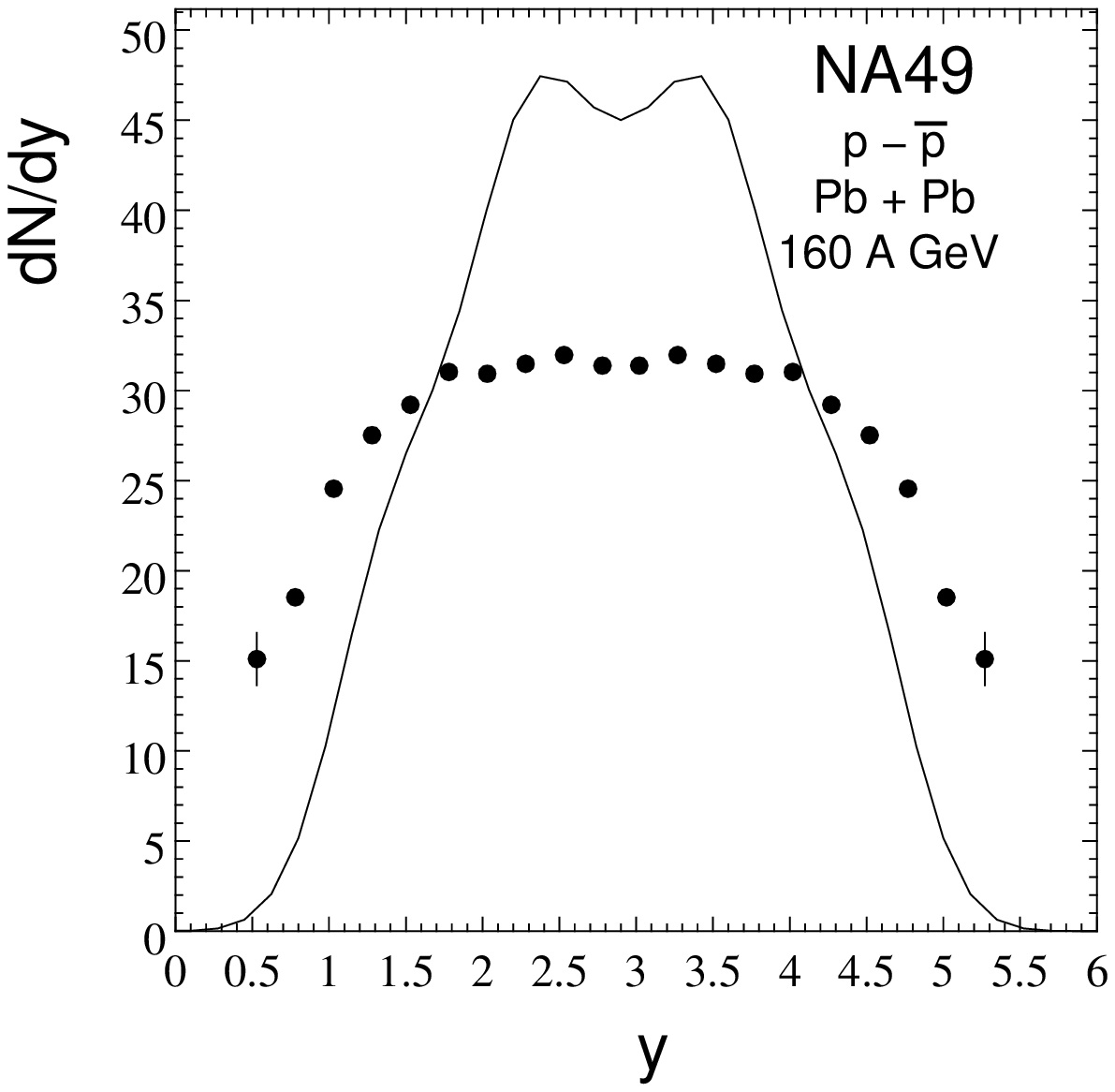}
         \hfill\\
   \end{minipage}\\

{\bf Figure \ref{Pblandau}:}
\end{center}

\newpage
\begin{center}
   \begin{minipage}[t]{\size}
         \epsfxsize \size \epsfbox{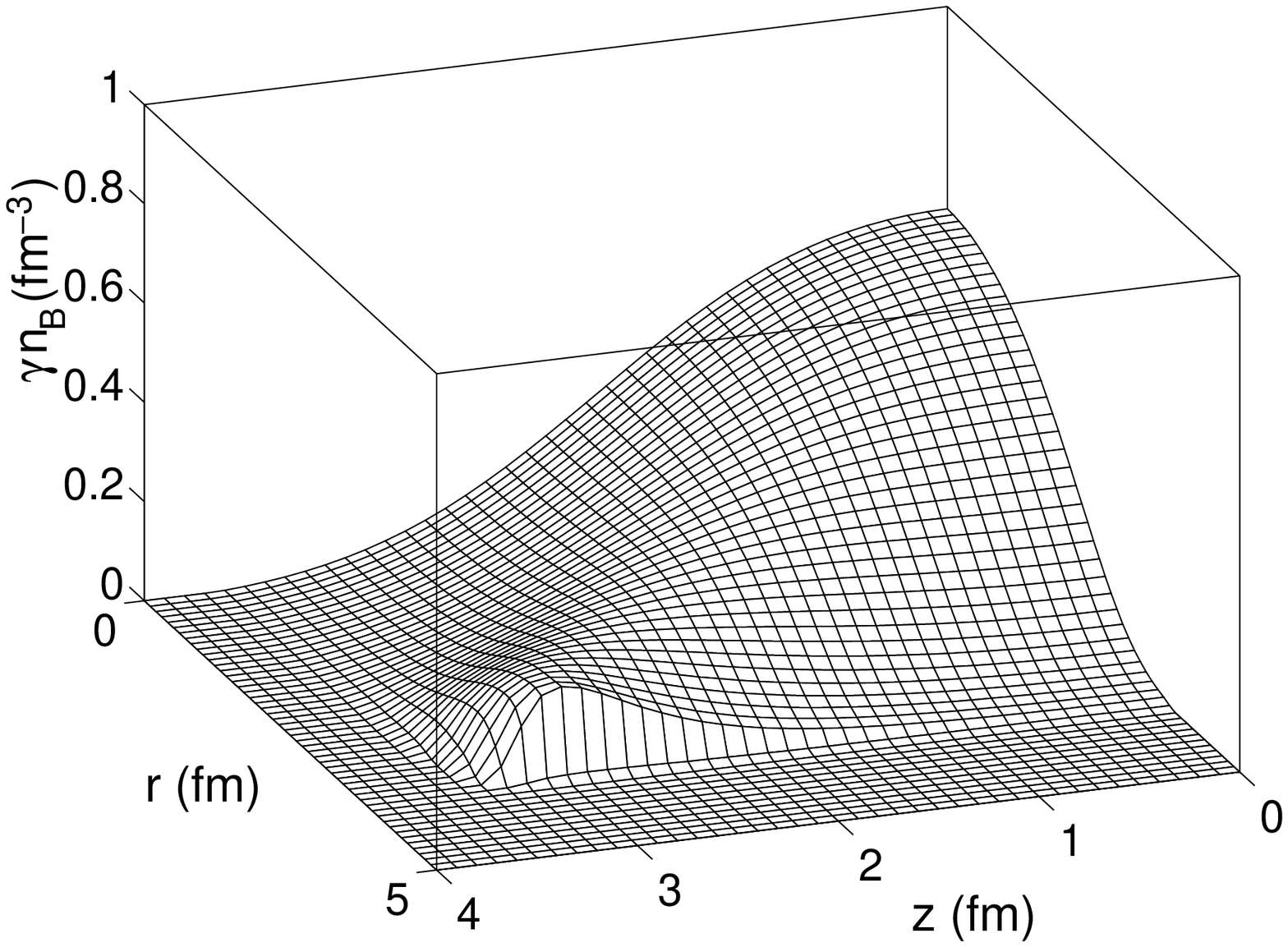}
         \hfill\\
   \end{minipage}\\

{\bf Figure \ref{inidens}:}
\end{center}

\newpage
   \vspace*{-2.0cm}
\begin{center}
   \begin{minipage}[t]{\sizer}
         \epsfxsize \sizer \epsfbox{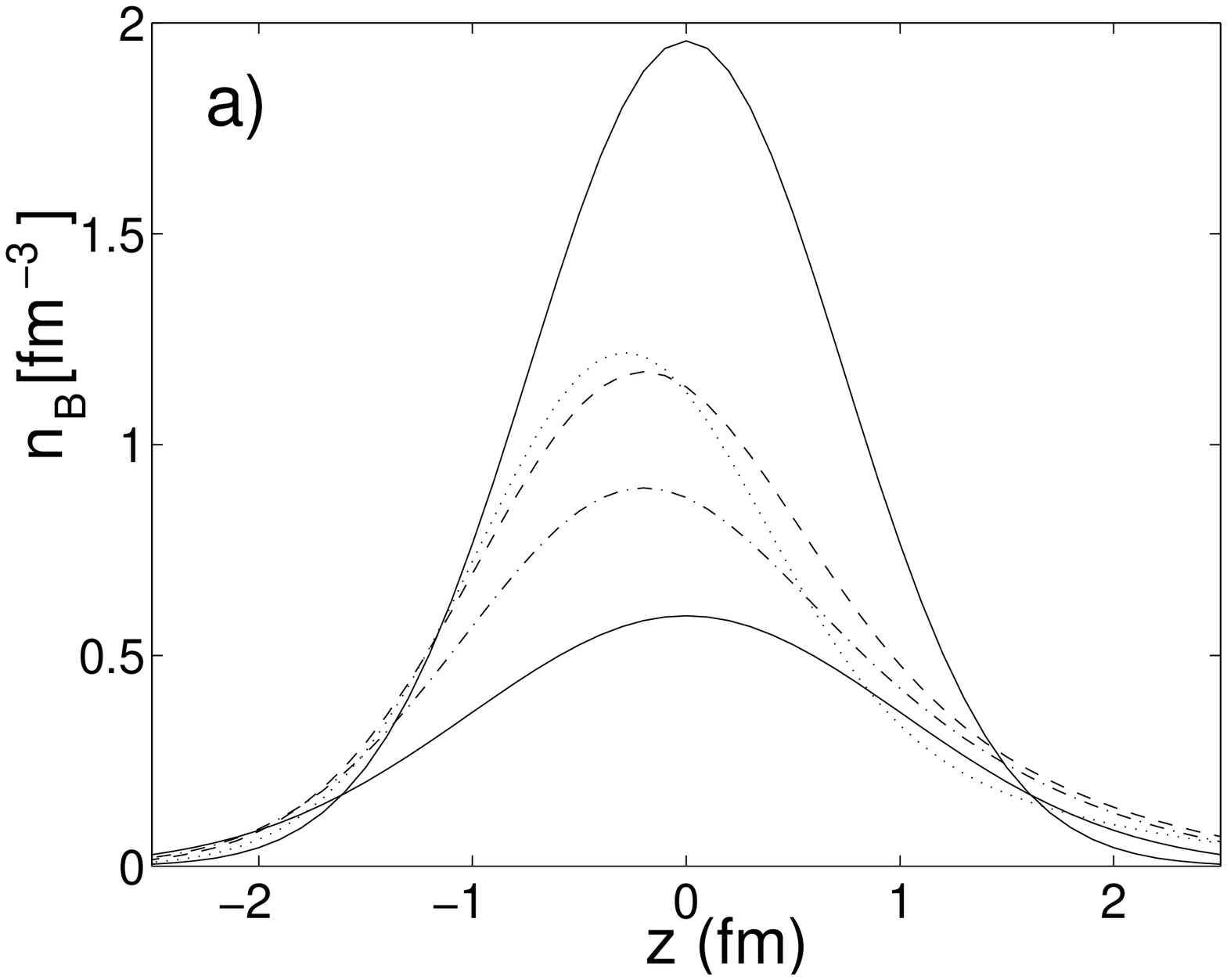}
         \hfill\\
   \end{minipage}

   \vspace*{-1.5truecm}
   \begin{minipage}[t]{\sizer}
         \epsfxsize \sizer \epsfbox{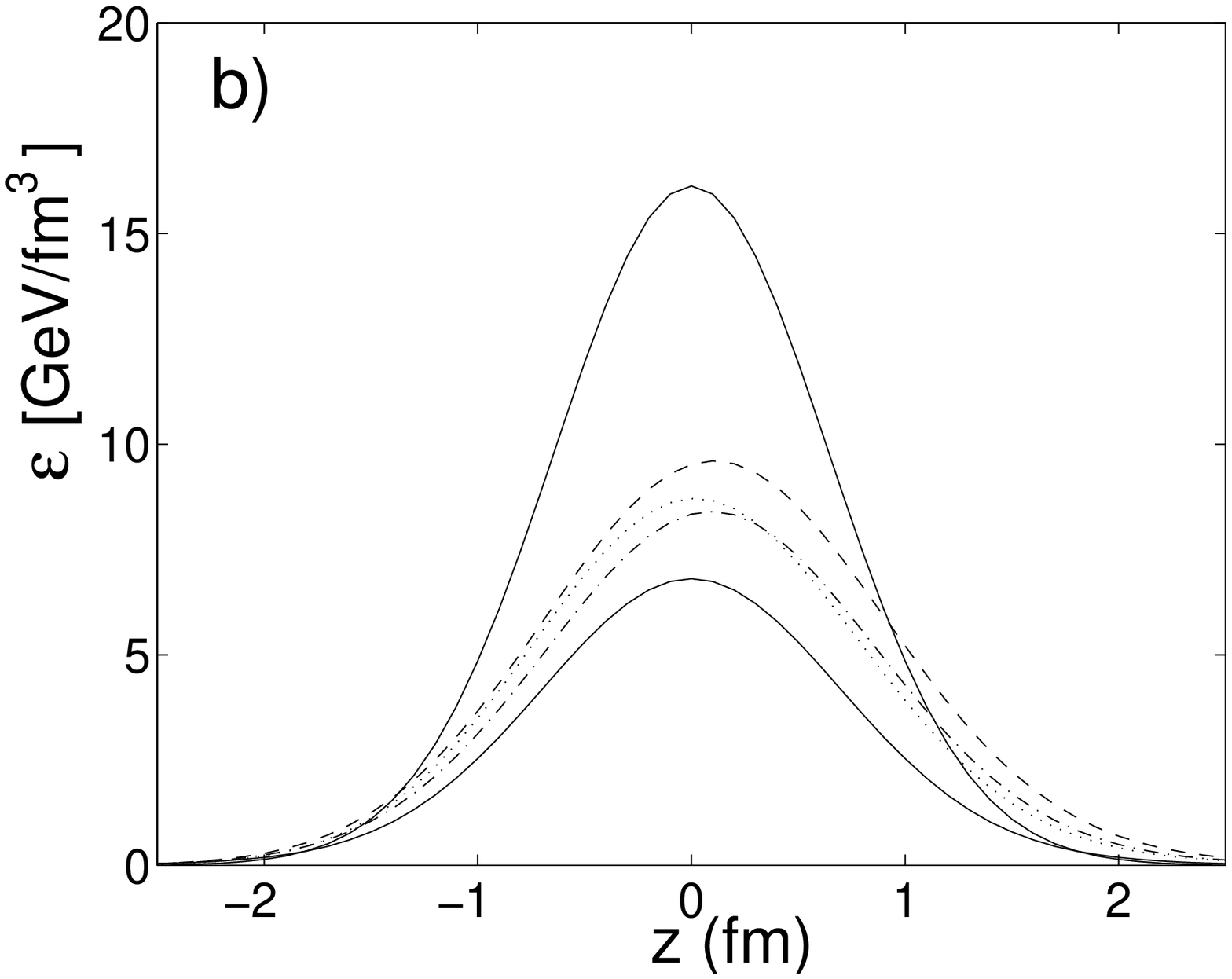}
         \hfill\\
   \end{minipage}\\

   \vspace*{-1.5truecm}
   \begin{minipage}[t]{\sizer}
         \epsfxsize \sizer \epsfbox{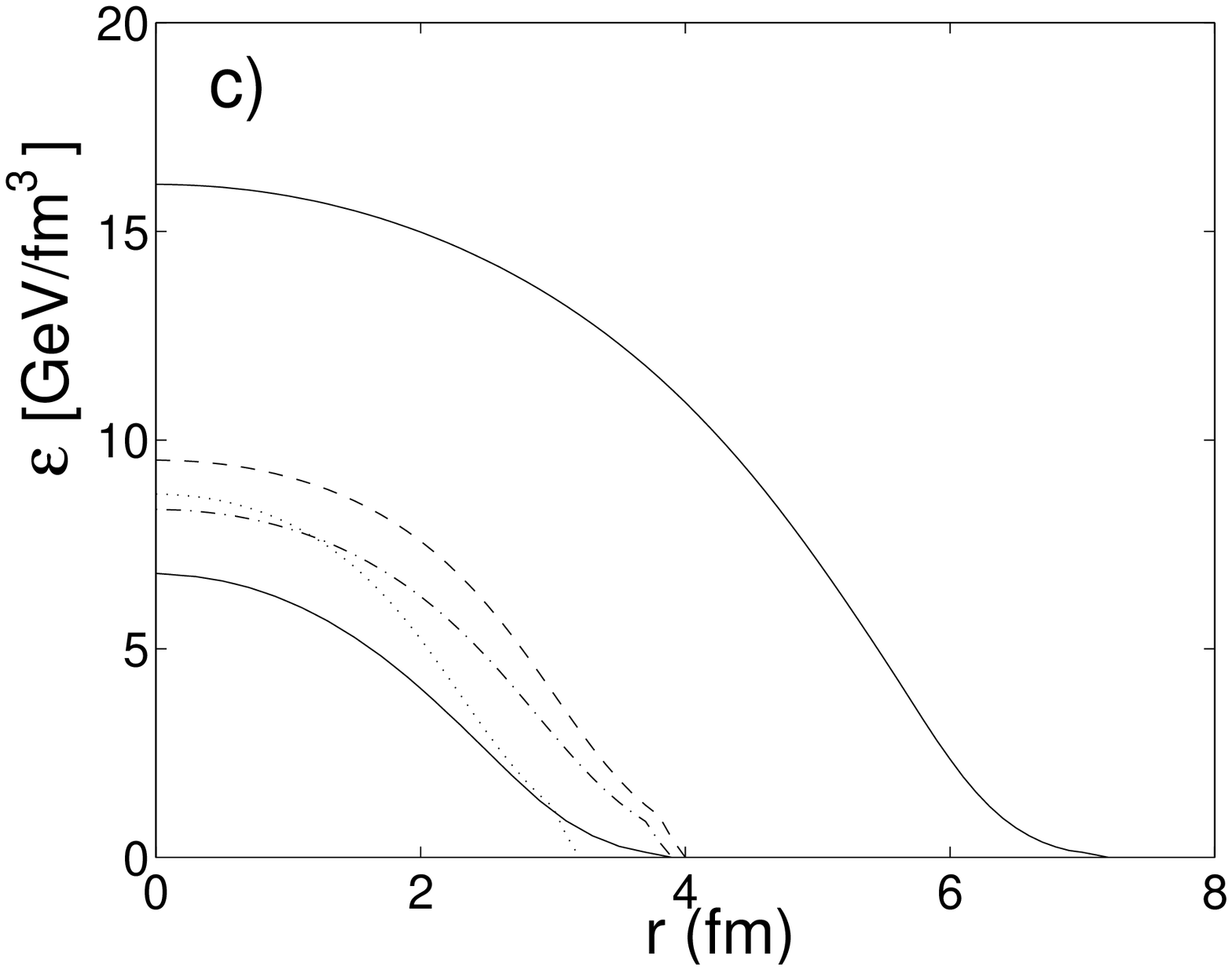}
         \hfill\\
   \end{minipage}

   \vspace*{-1cm}
   {\bf Figure \ref{inipro}:}
\end{center}

\newpage
\begin{center}
   \begin{minipage}[t]{\size}
         \epsfxsize \size \epsfbox{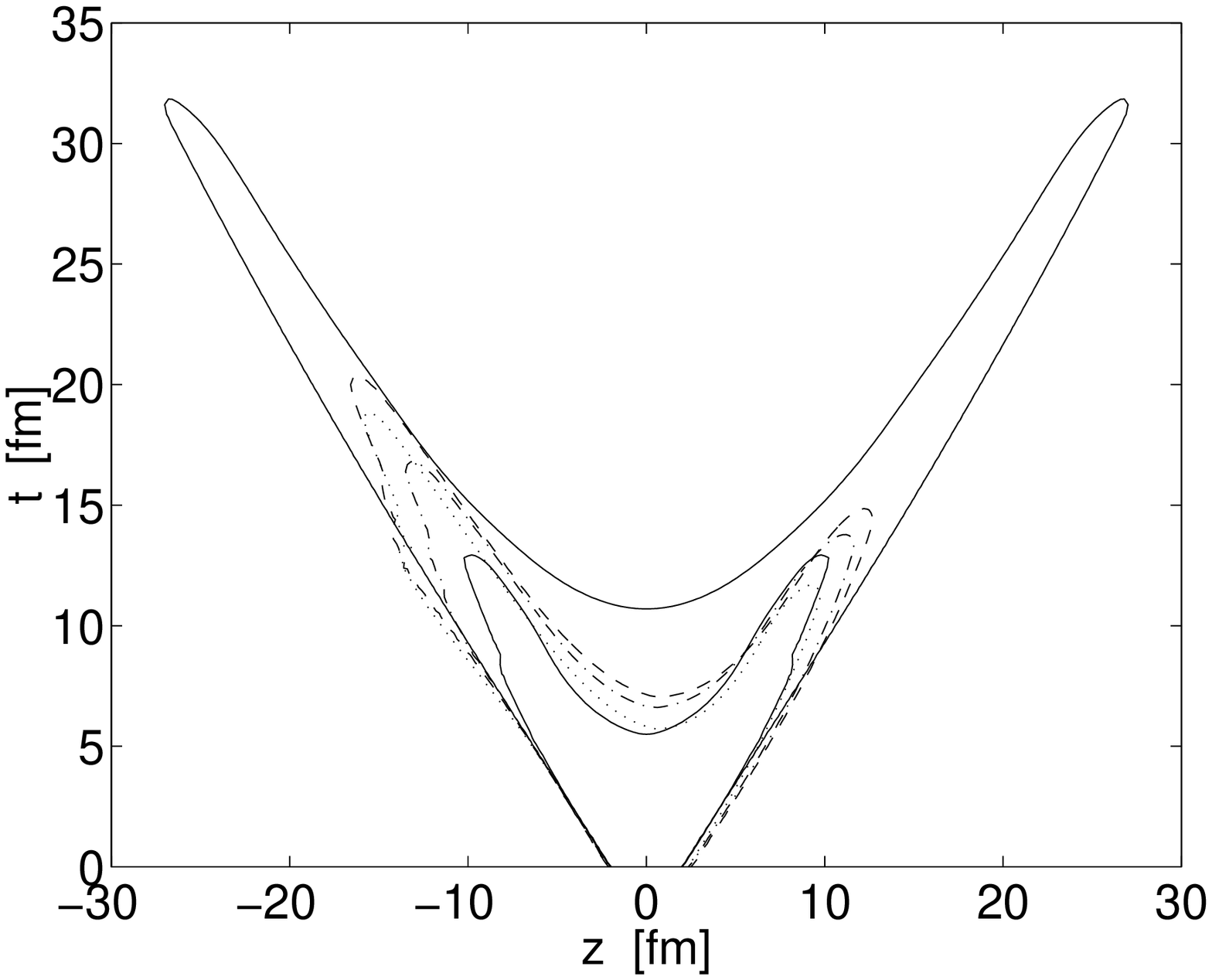}
         \hfill\\
   \end{minipage}\\

{\bf Figure \ref{freeze}:}
\end{center}

\newpage
\begin{center}
  \begin{minipage}[t]{\sizen}
        \epsfxsize \sizen \epsfbox{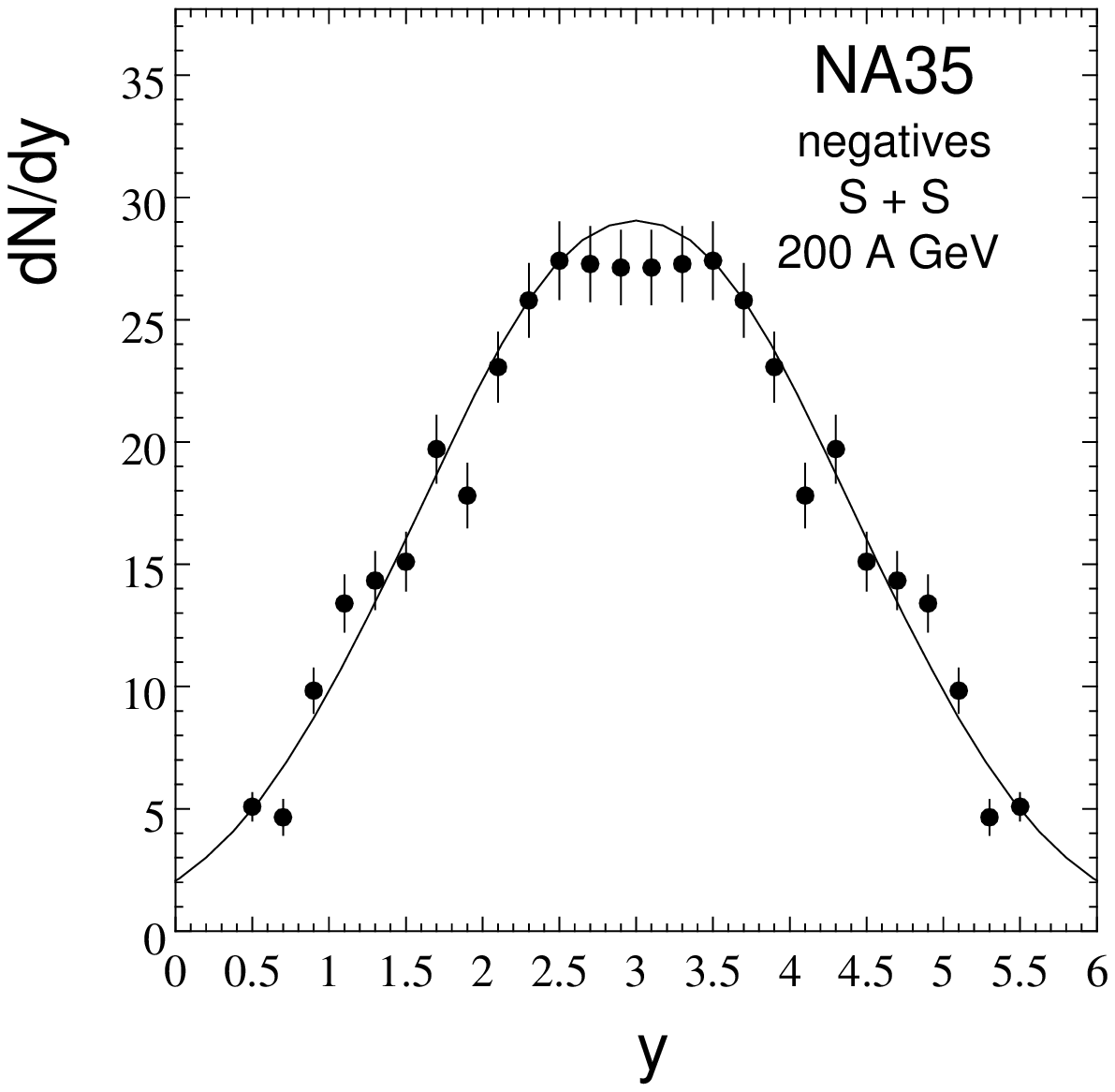}
        \hfill
  \end{minipage}
  \hfill
  \begin{minipage}[t]{\sizen}
        \epsfxsize \sizen \epsfbox{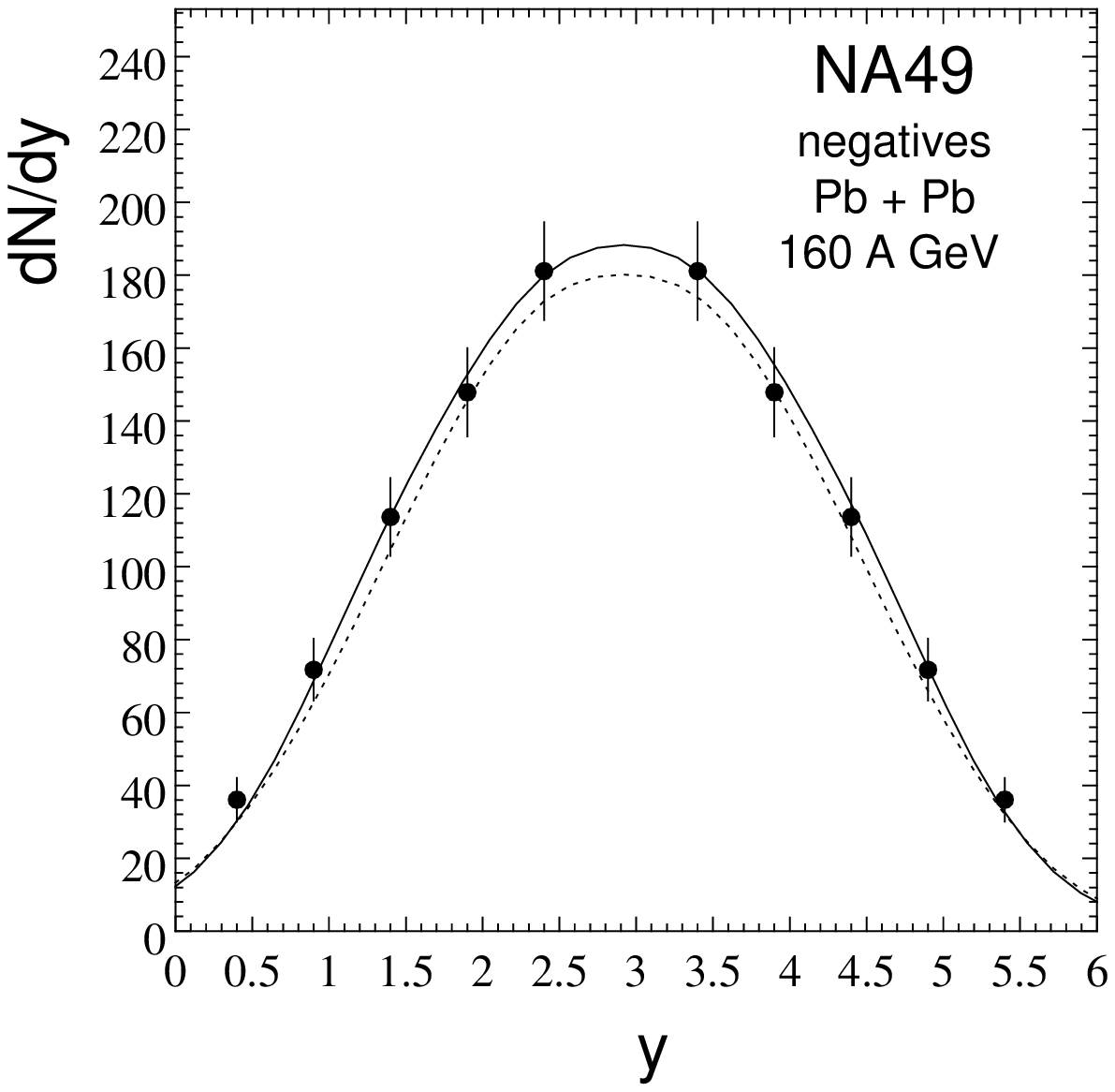}
        \hfill
  \end{minipage}

  \vspace*{-1.5truecm}
  \begin{minipage}[t]{\sizen}
        \epsfxsize \sizen \epsfbox{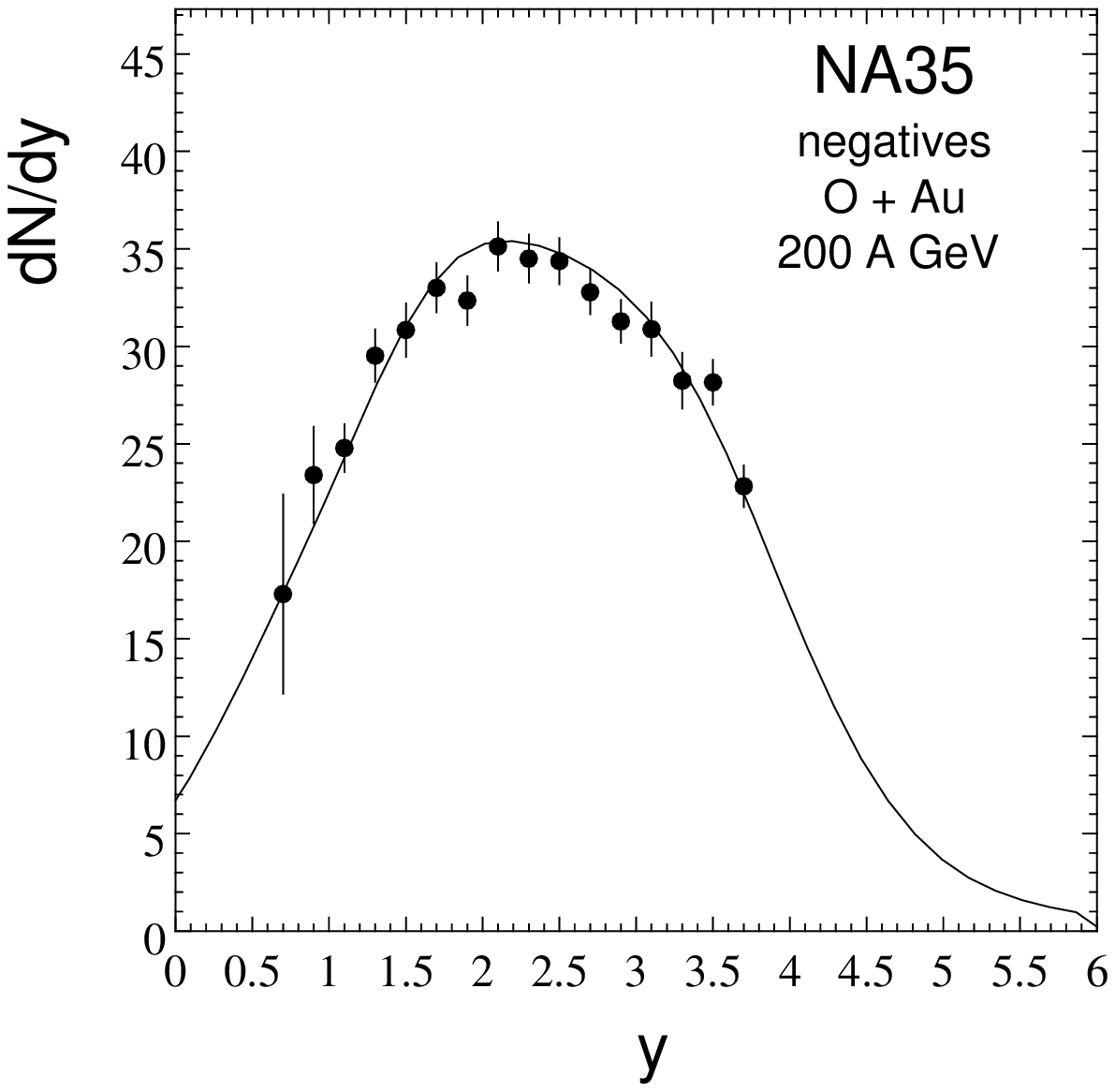}
        \hfill
  \end{minipage}
  \hfill
  \begin{minipage}[t]{\sizen}
        \epsfxsize \sizen \epsfbox{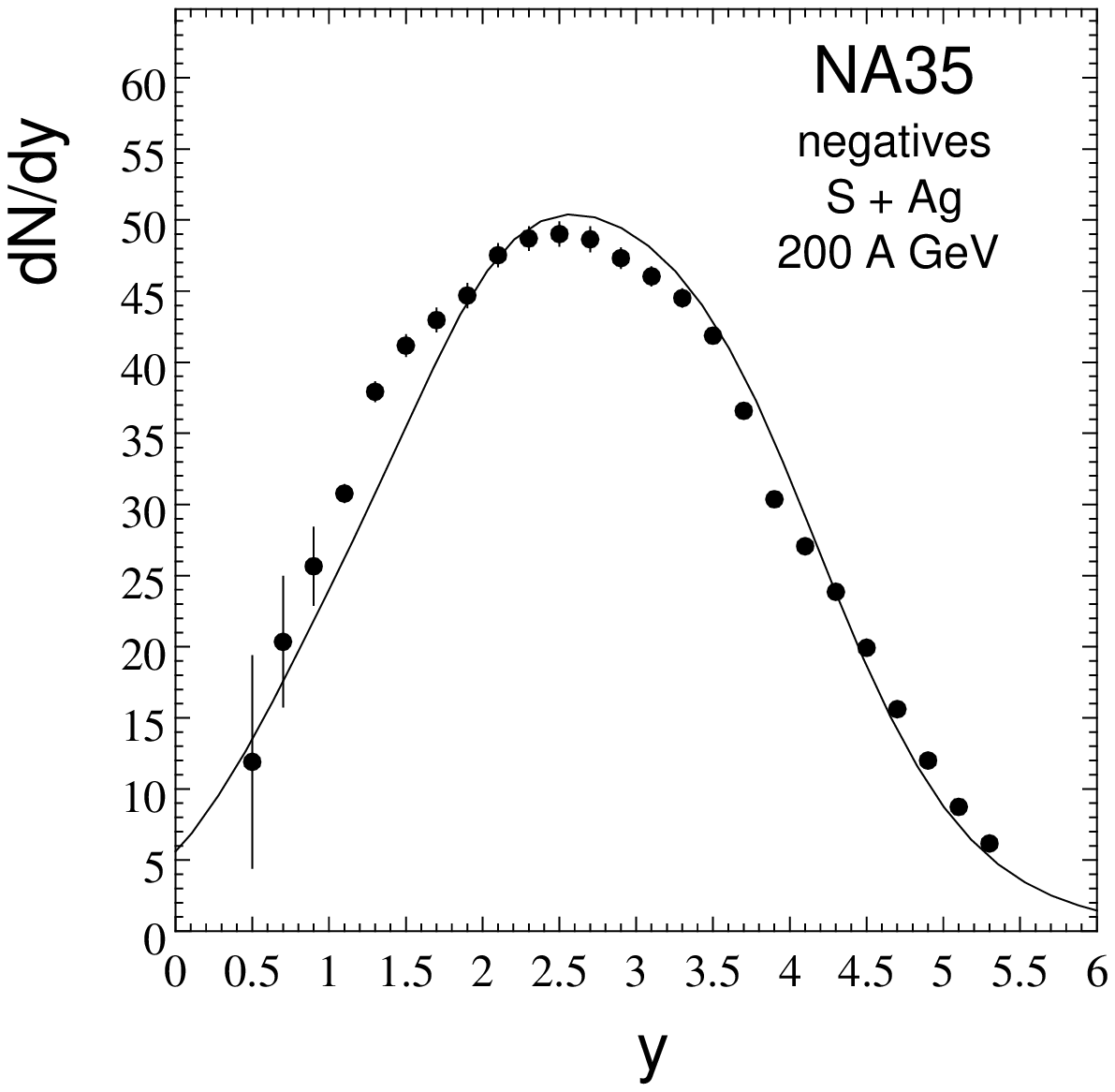}
        \hfill
  \end{minipage}

  \vspace*{-1.5truecm}
  \begin{minipage}[t]{\sizen}
        \epsfxsize \sizen \epsfbox{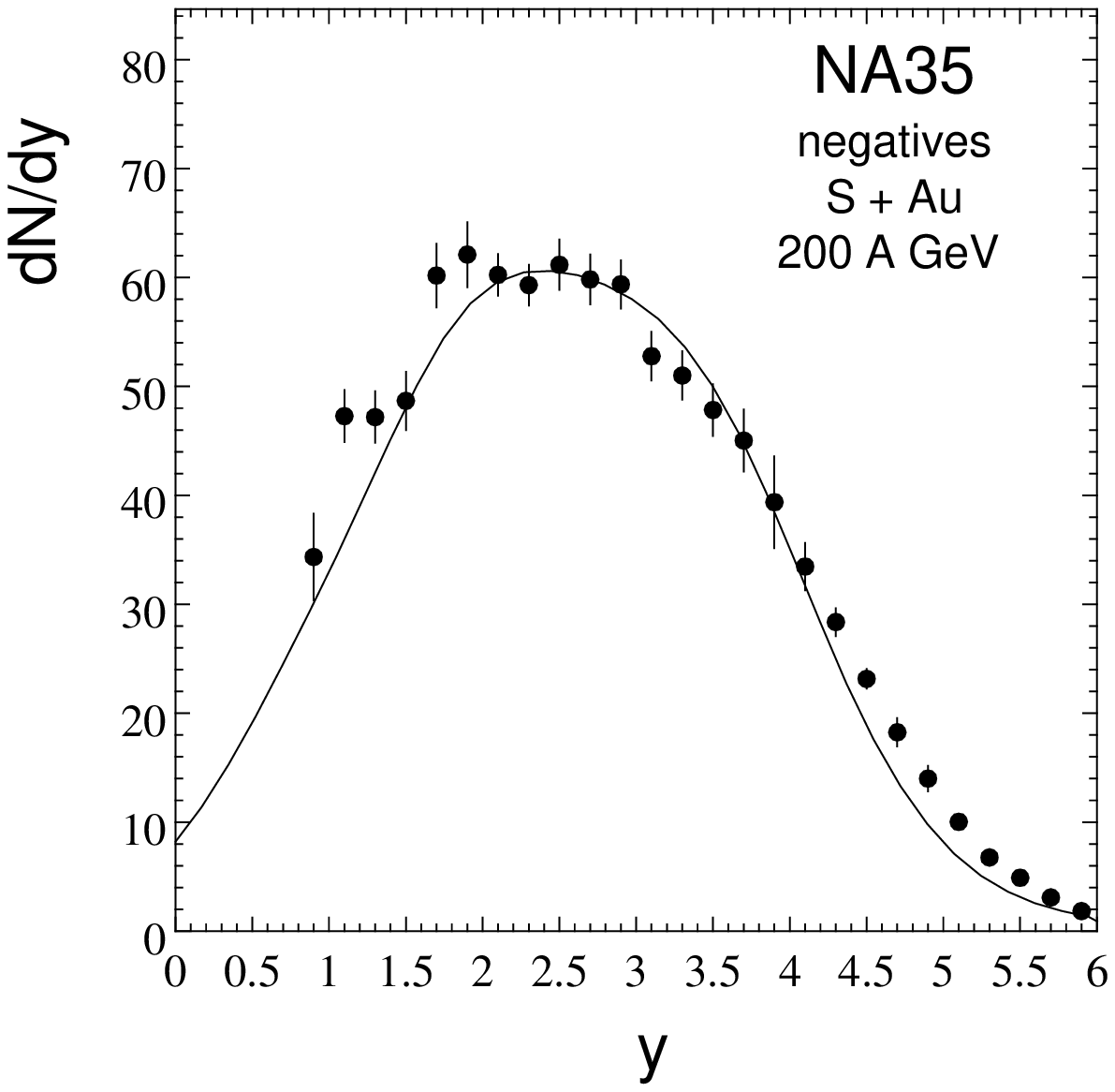}
        \hfill
  \end{minipage}
  \hfill
  \begin{minipage}[t]{\sizen}
        \epsfxsize \sizen \epsfbox{dummy.ps}
        \hfill
  \end{minipage}\\
  {\bf Figure \ref{negrap}:}
\end{center}

\newpage
\begin{center}
  \begin{minipage}[t]{\sizen}
        \epsfxsize \sizen \epsfbox{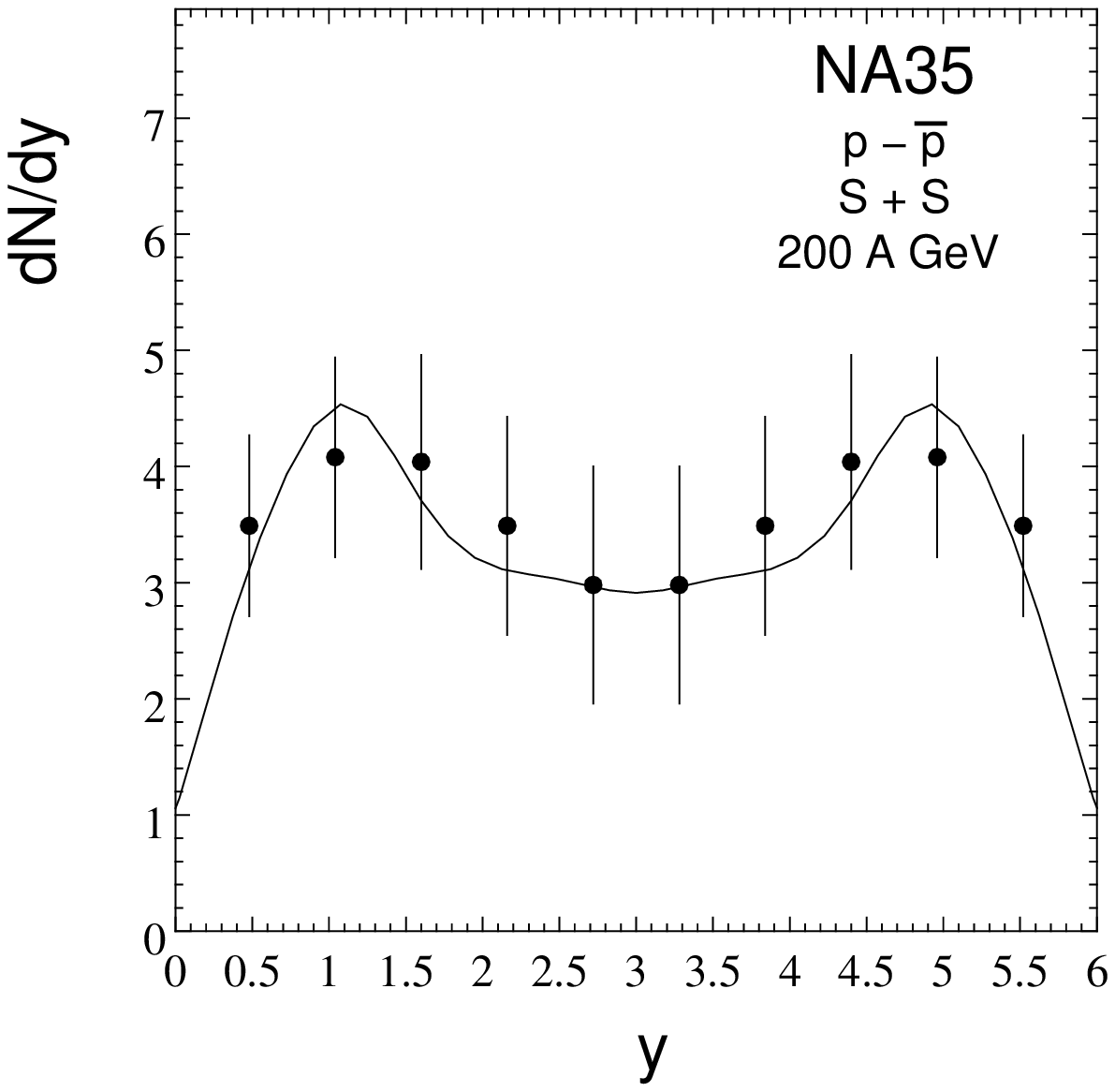}
        \hfill
  \end{minipage}
  \hfill
  \begin{minipage}[t]{\sizen}
        \epsfxsize \sizen \epsfbox{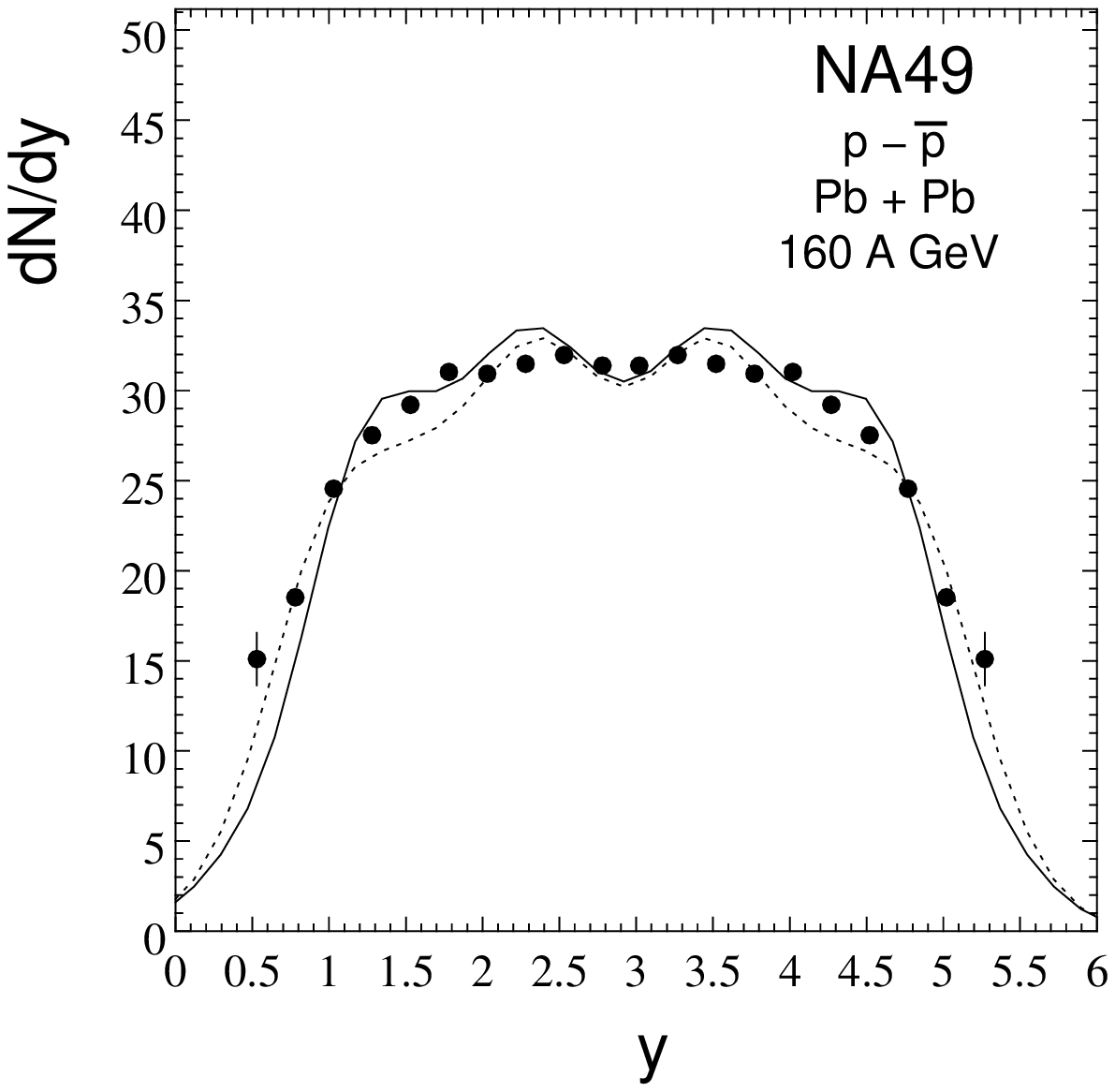}
        \hfill
  \end{minipage}

  \vspace*{-1.5truecm}
  \begin{minipage}[t]{\sizen}
        \epsfxsize \sizen \epsfbox{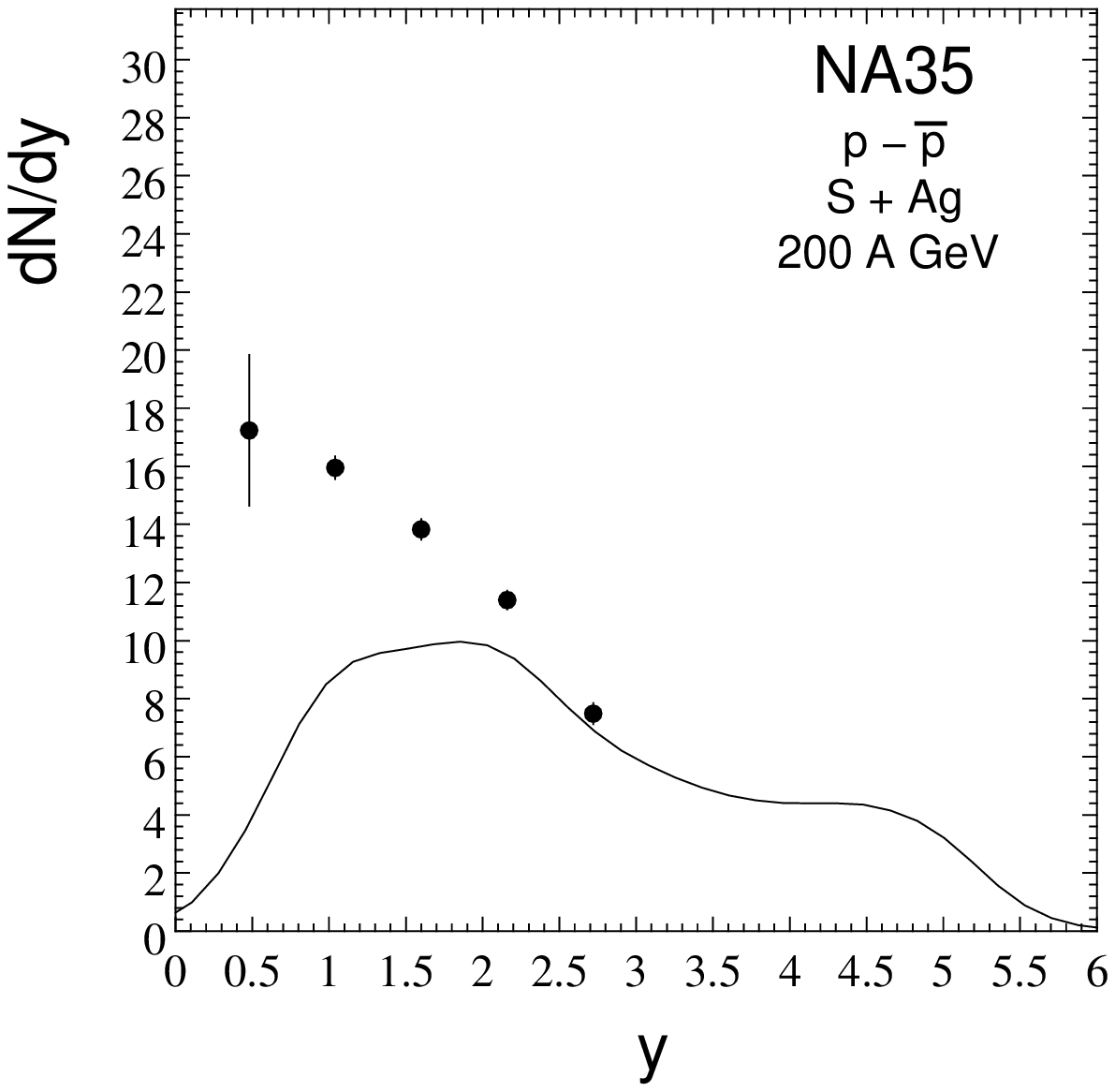}
        \hfill
  \end{minipage}
  \hfill
  \begin{minipage}[t]{\sizen}
        \epsfxsize \sizen \epsfbox{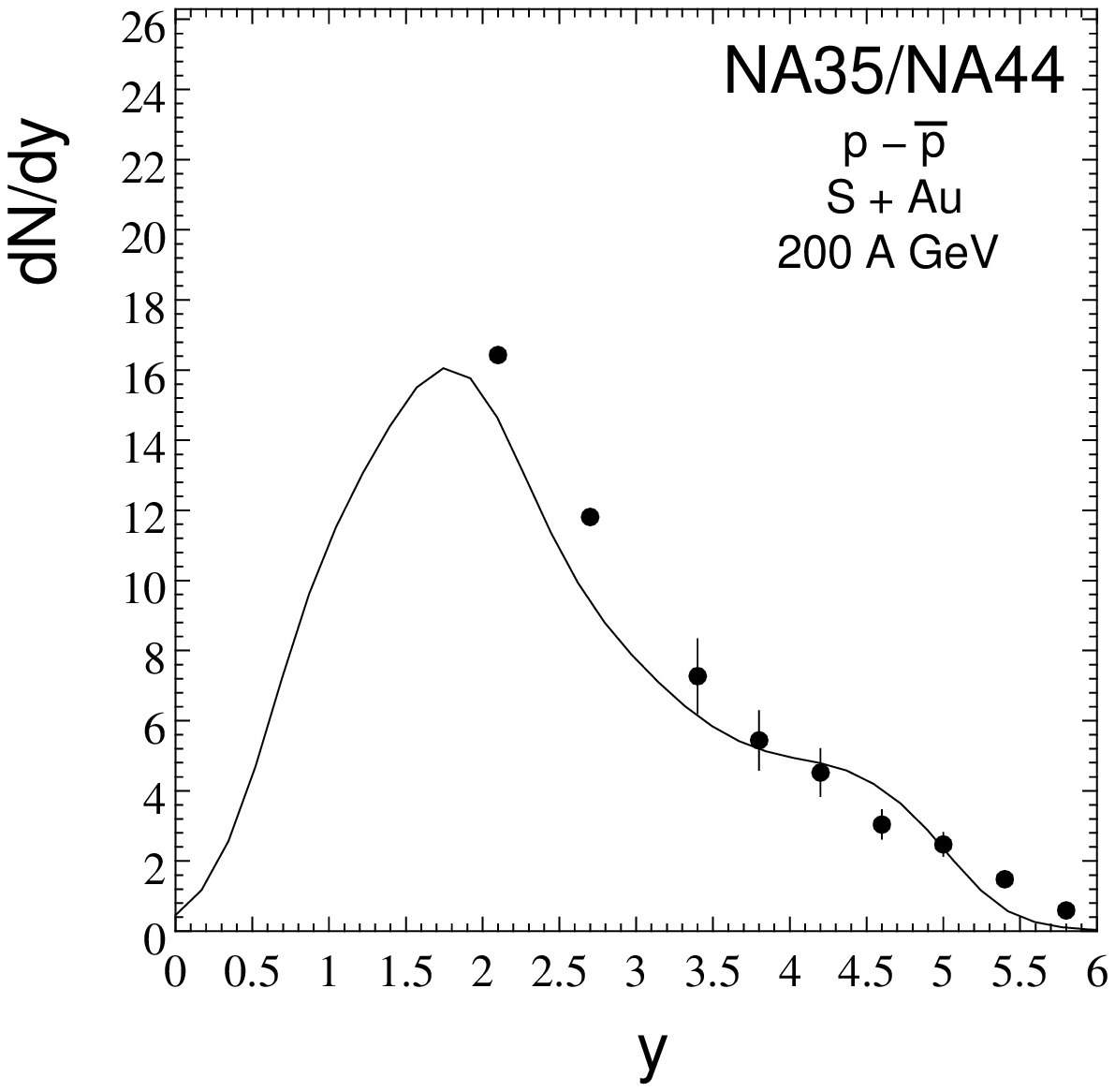}
        \hfill
  \end{minipage}

  \vspace*{-1.5truecm}
  \begin{minipage}[t]{\sizen}
        \epsfxsize \sizen \epsfbox{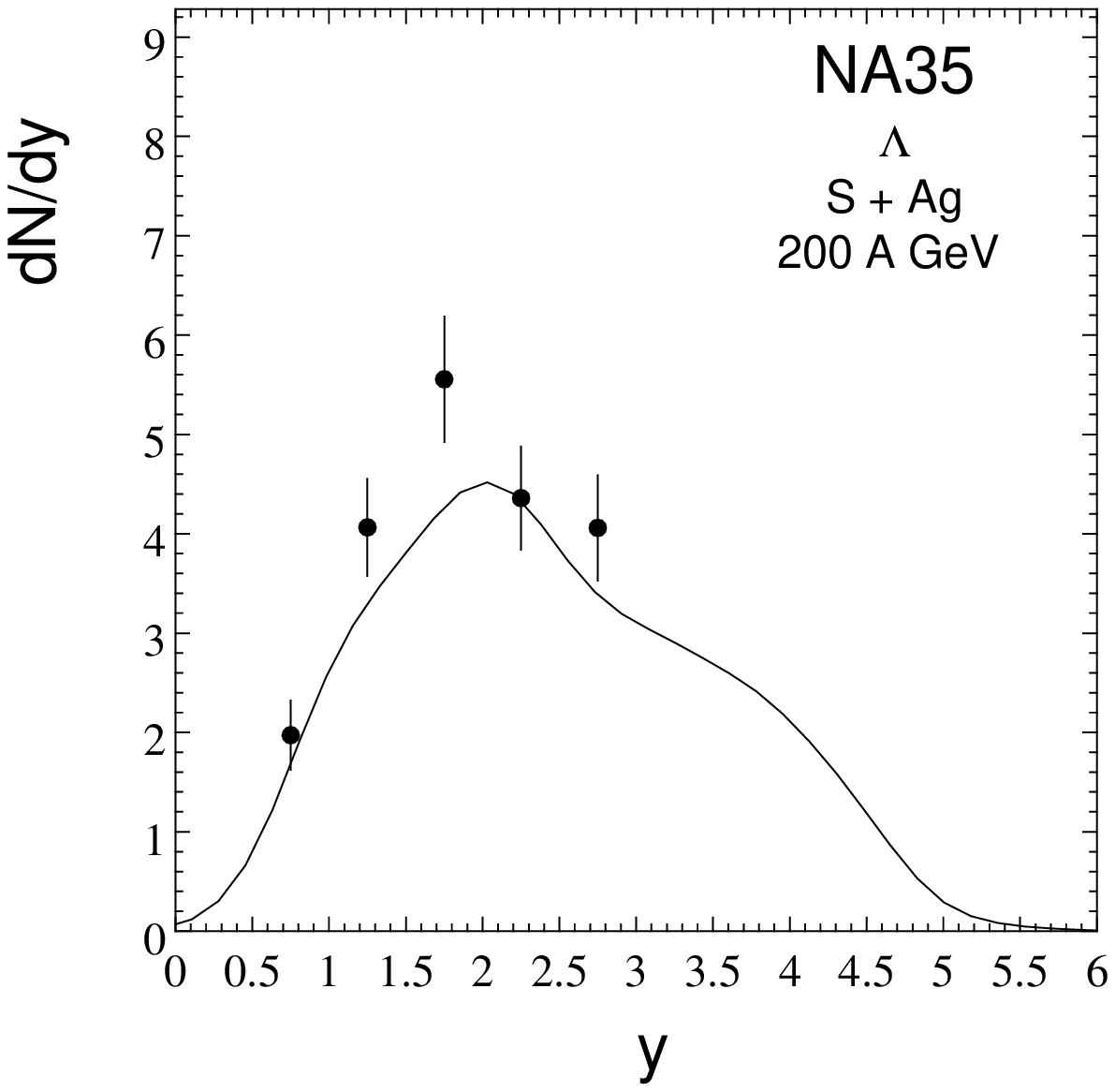}
        \hfill
  \end{minipage}
  \hfill
  \begin{minipage}[t]{\sizen}
        \epsfxsize \sizen \epsfbox{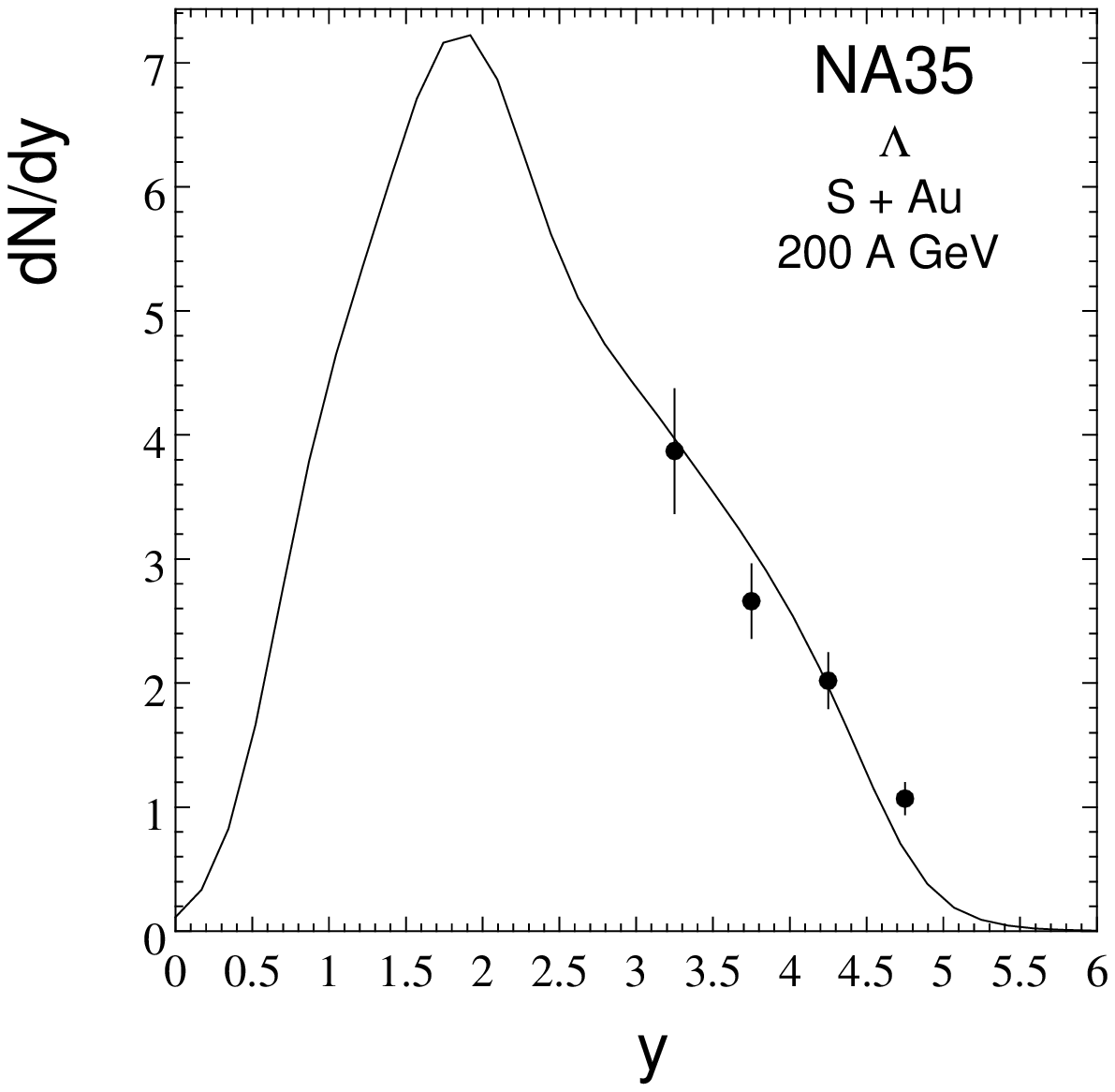}
        \hfill
  \end{minipage}

  {\bf Figure \ref{nprorap}:}
\end{center}

\newpage

\begin{center}
  \begin{minipage}[t]{\sizen}
        \epsfxsize \sizen \epsfbox{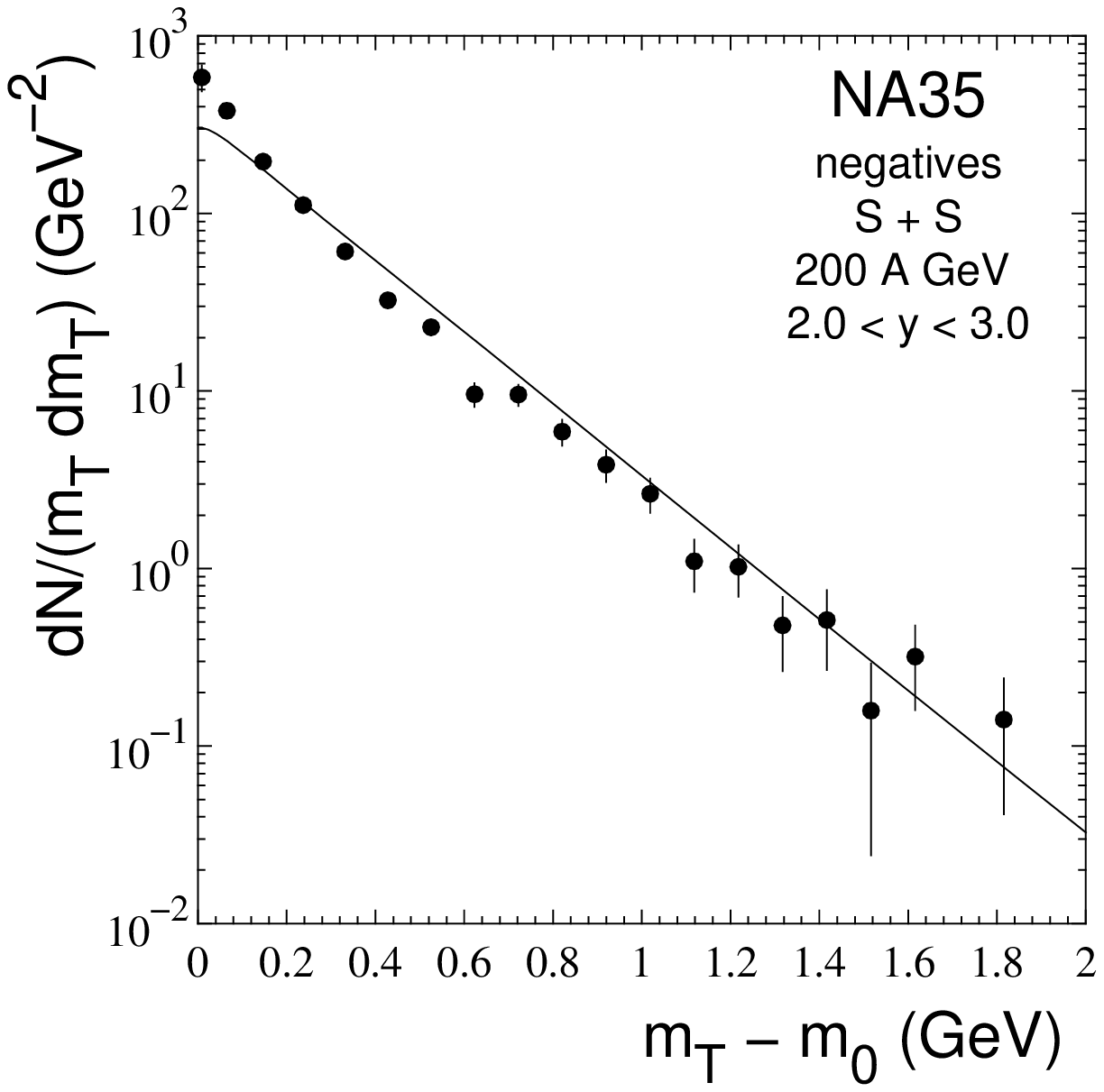}
        \hfill
  \end{minipage}
  \hfill
  \begin{minipage}[t]{\sizen}
        \epsfxsize \sizen \epsfbox{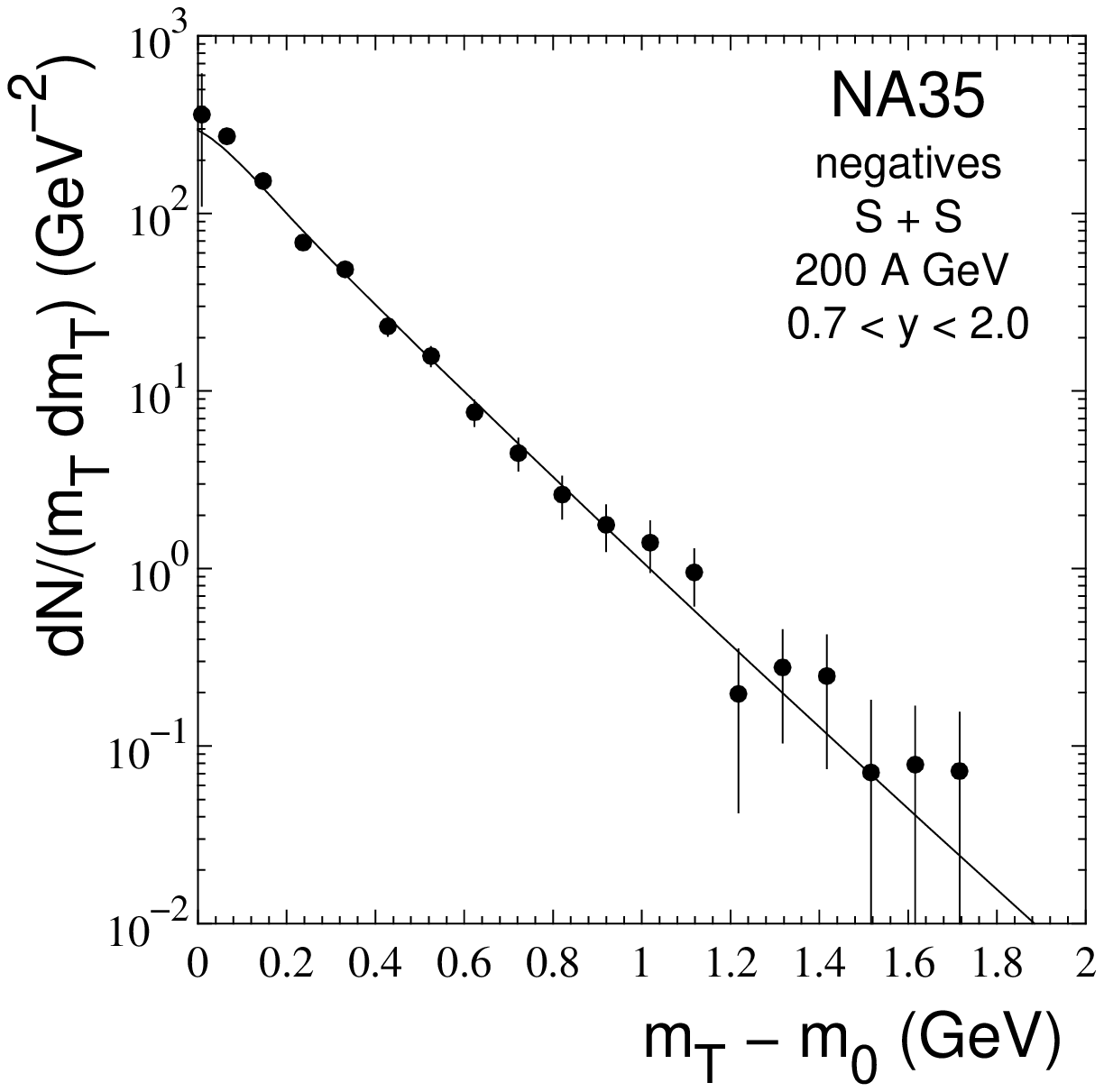}
        \hfill
  \end{minipage}\\

  \vspace*{-1.5truecm}
  \begin{minipage}[t]{\sizen}
        \epsfxsize \sizen \epsfbox{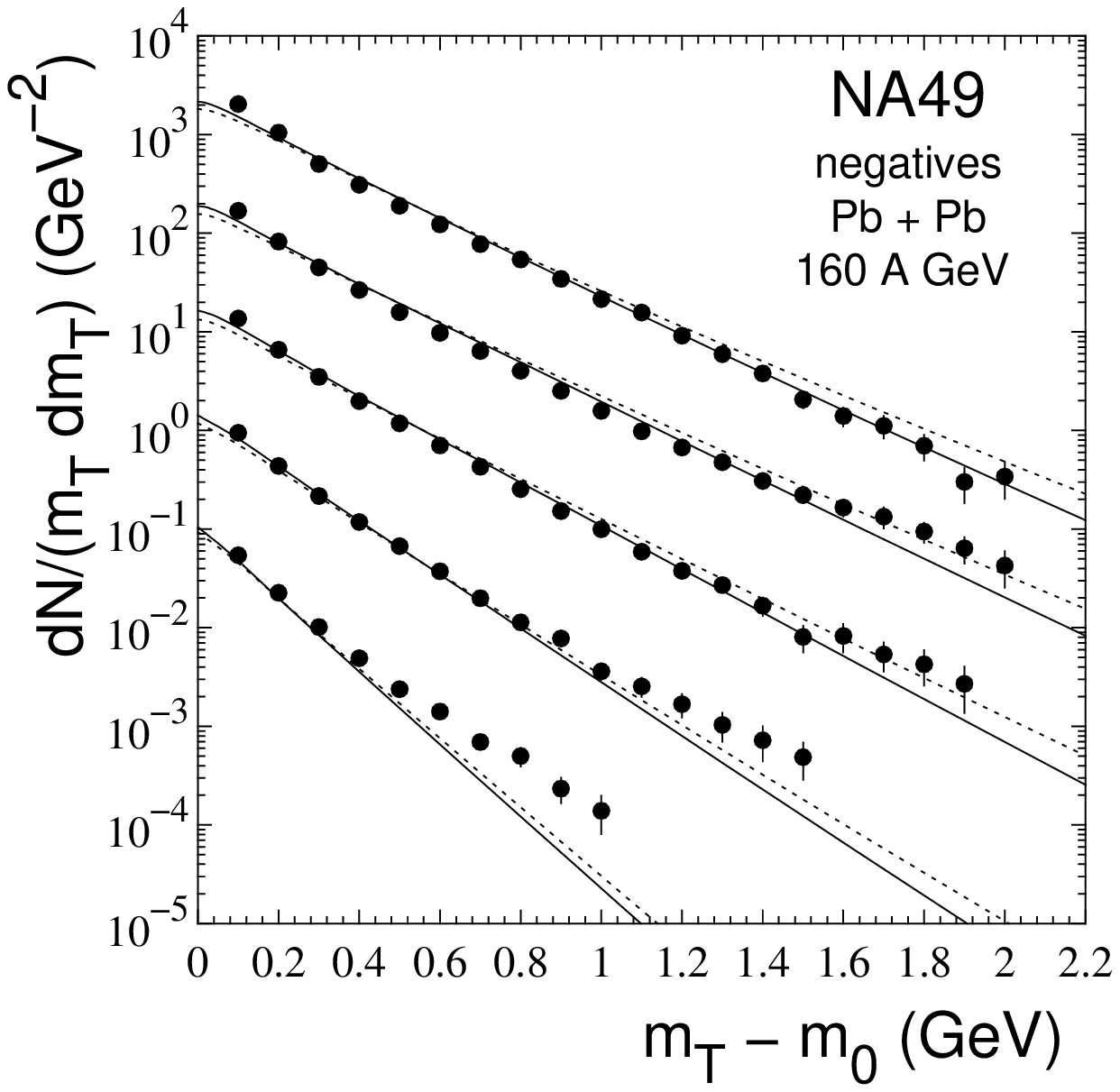}
        \hfill
  \end{minipage}
  \hfill
  \begin{minipage}[t]{\sizen}
        \epsfxsize \sizen \epsfbox{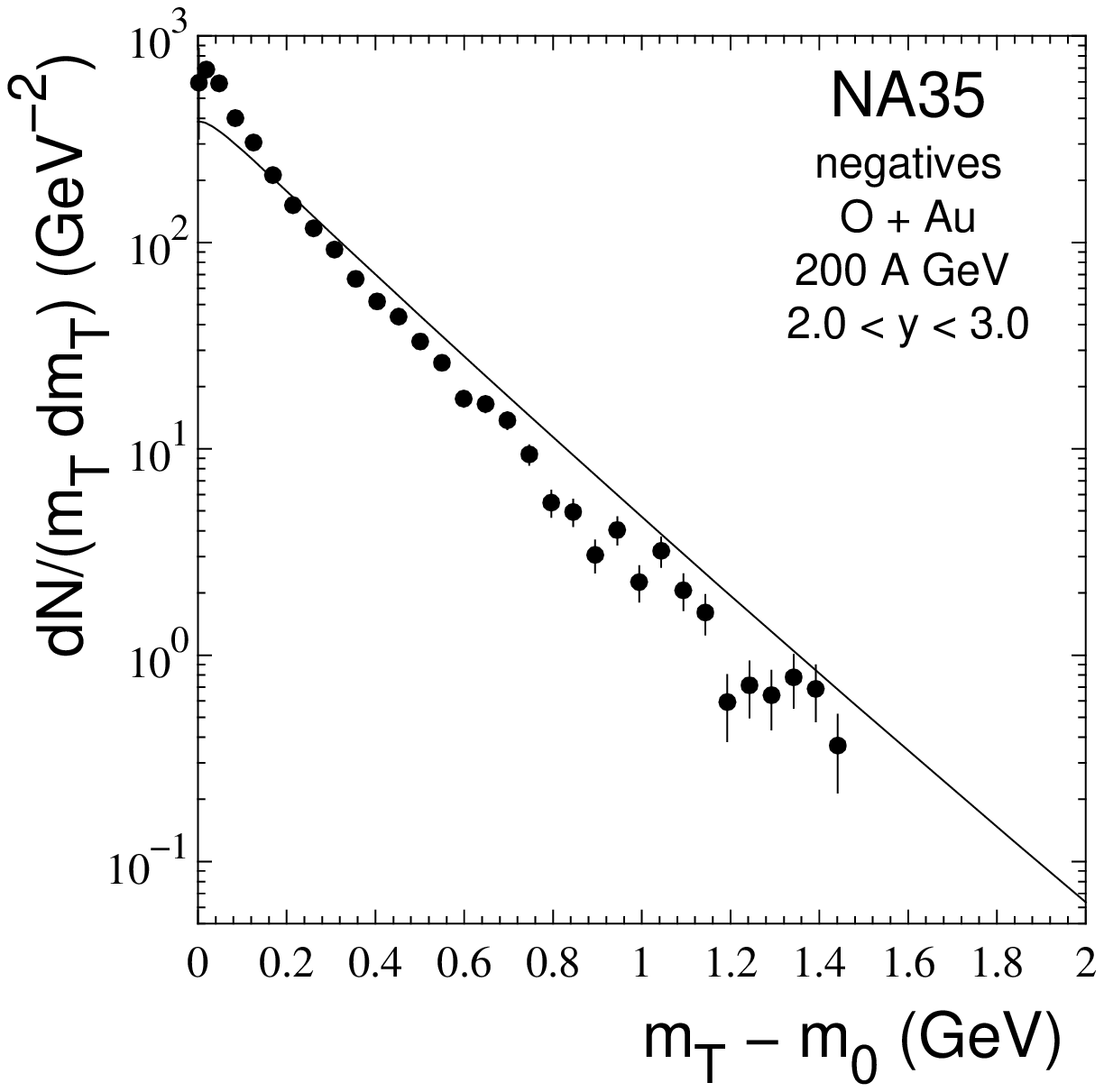}
  \end{minipage}

  \vspace*{-1.5truecm}
  \begin{minipage}[t]{\sizen}
        \epsfxsize \sizen \epsfbox{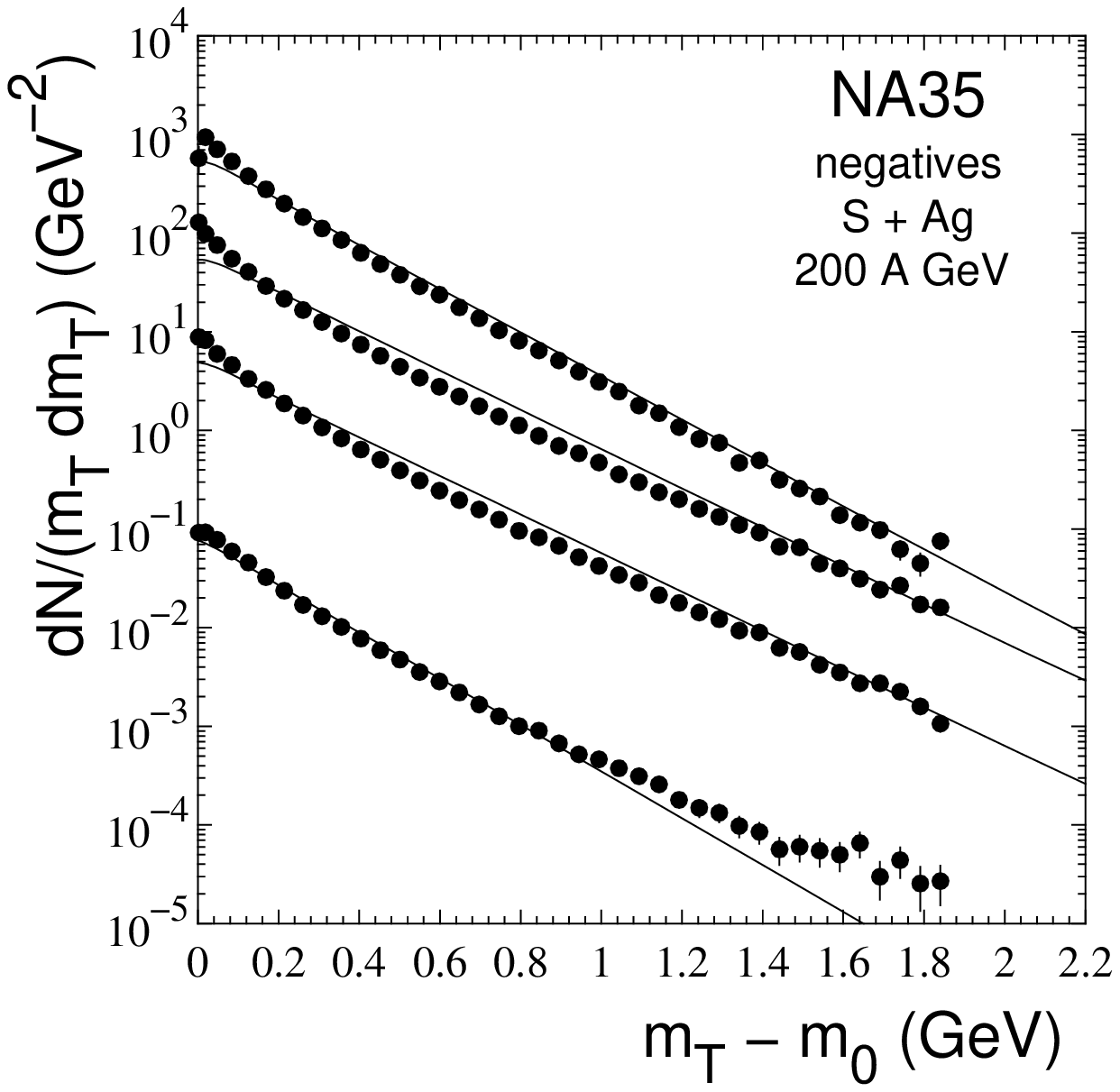}
        \hfill
  \end{minipage}
  \hfill
  \begin{minipage}[t]{\sizen}
        \epsfxsize \sizen \epsfbox{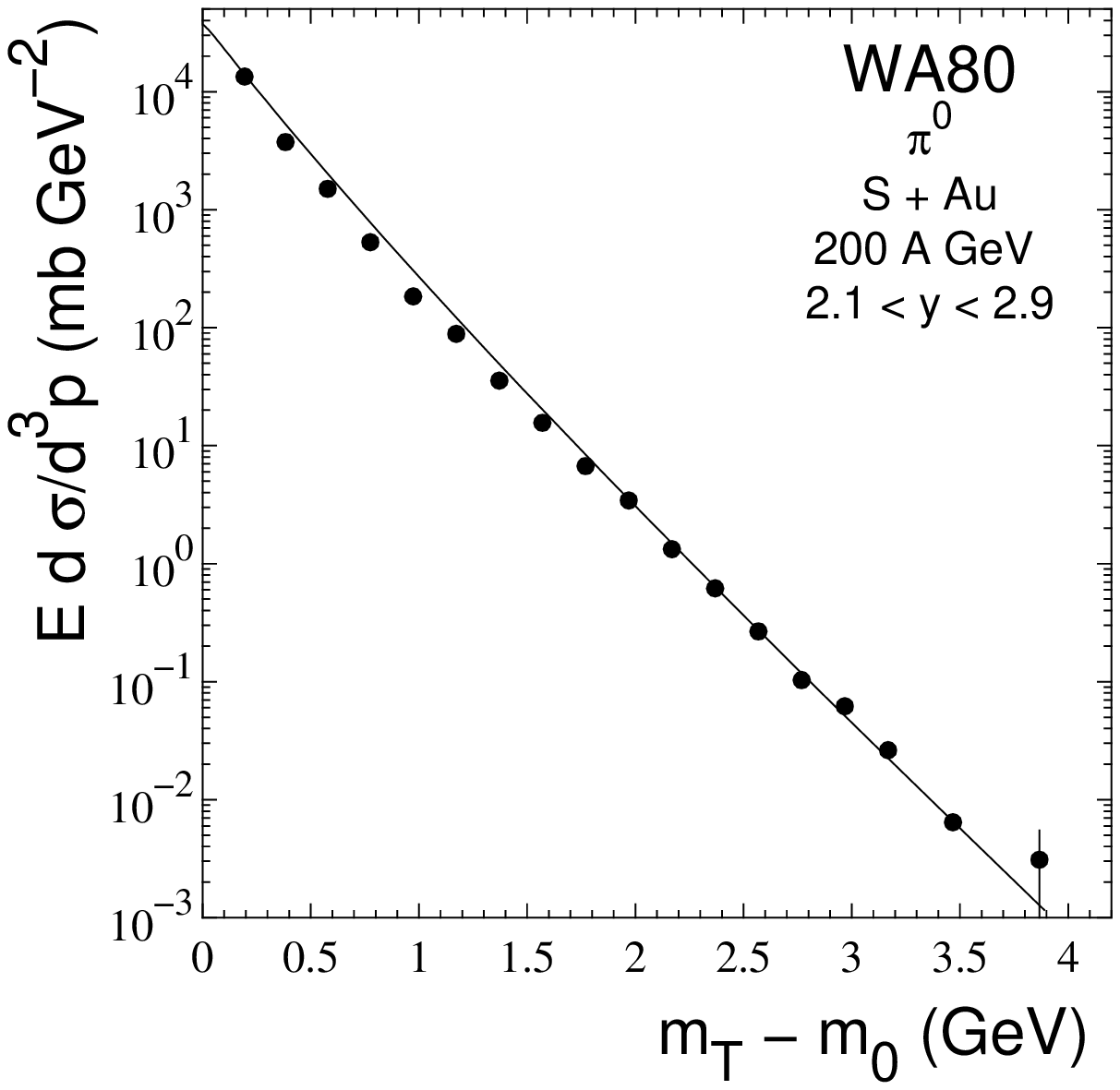}
        \hfill
  \end{minipage}
  {\bf Figure \ref{negpt}:}
\end{center}

\newpage
\begin{center}
  \begin{minipage}[t]{\sizen}
        \epsfxsize \sizen \epsfbox{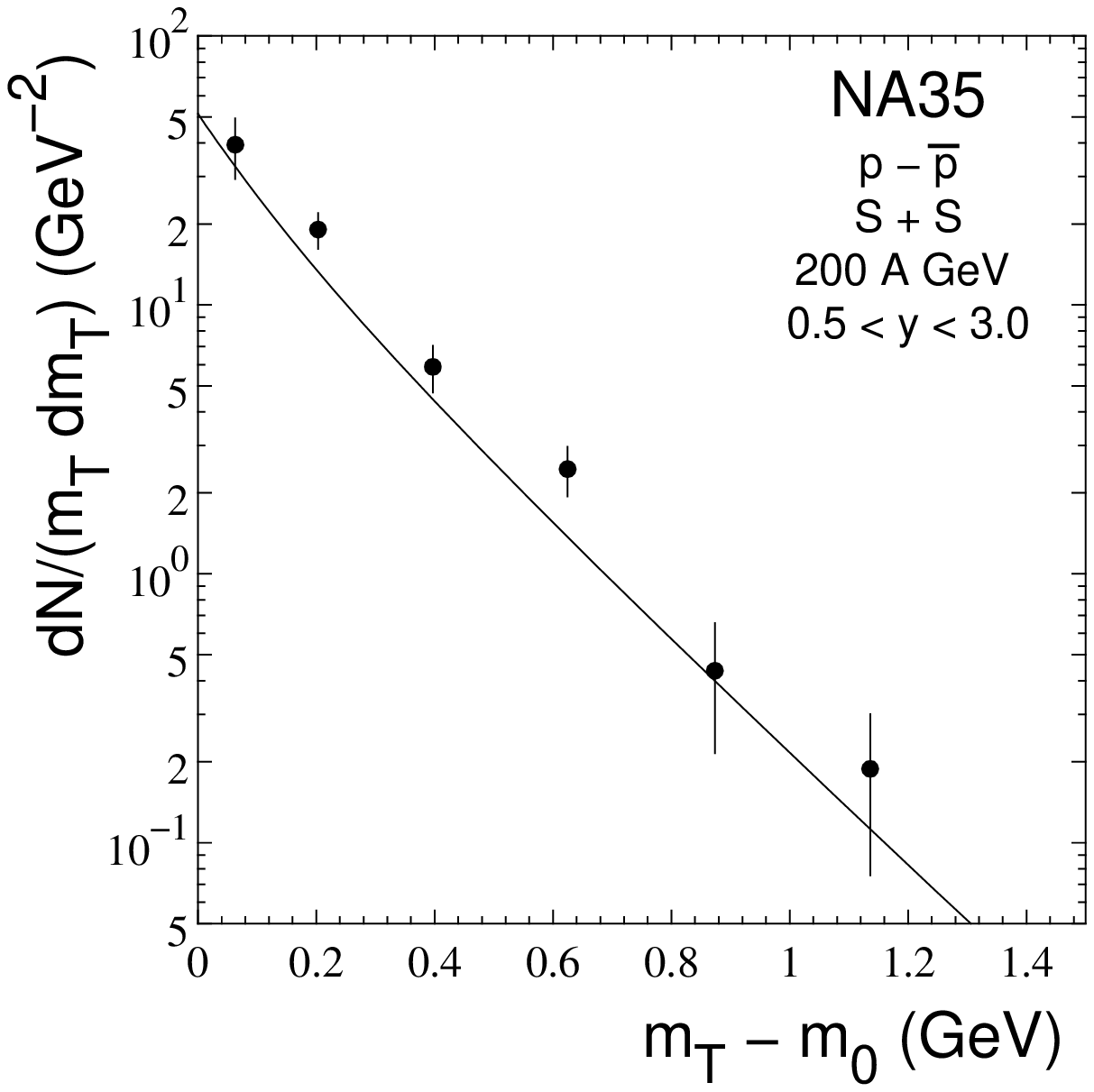}
        \hfill
  \end{minipage}
  \hfill
  \begin{minipage}[t]{\sizen}
        \epsfxsize \sizen \epsfbox{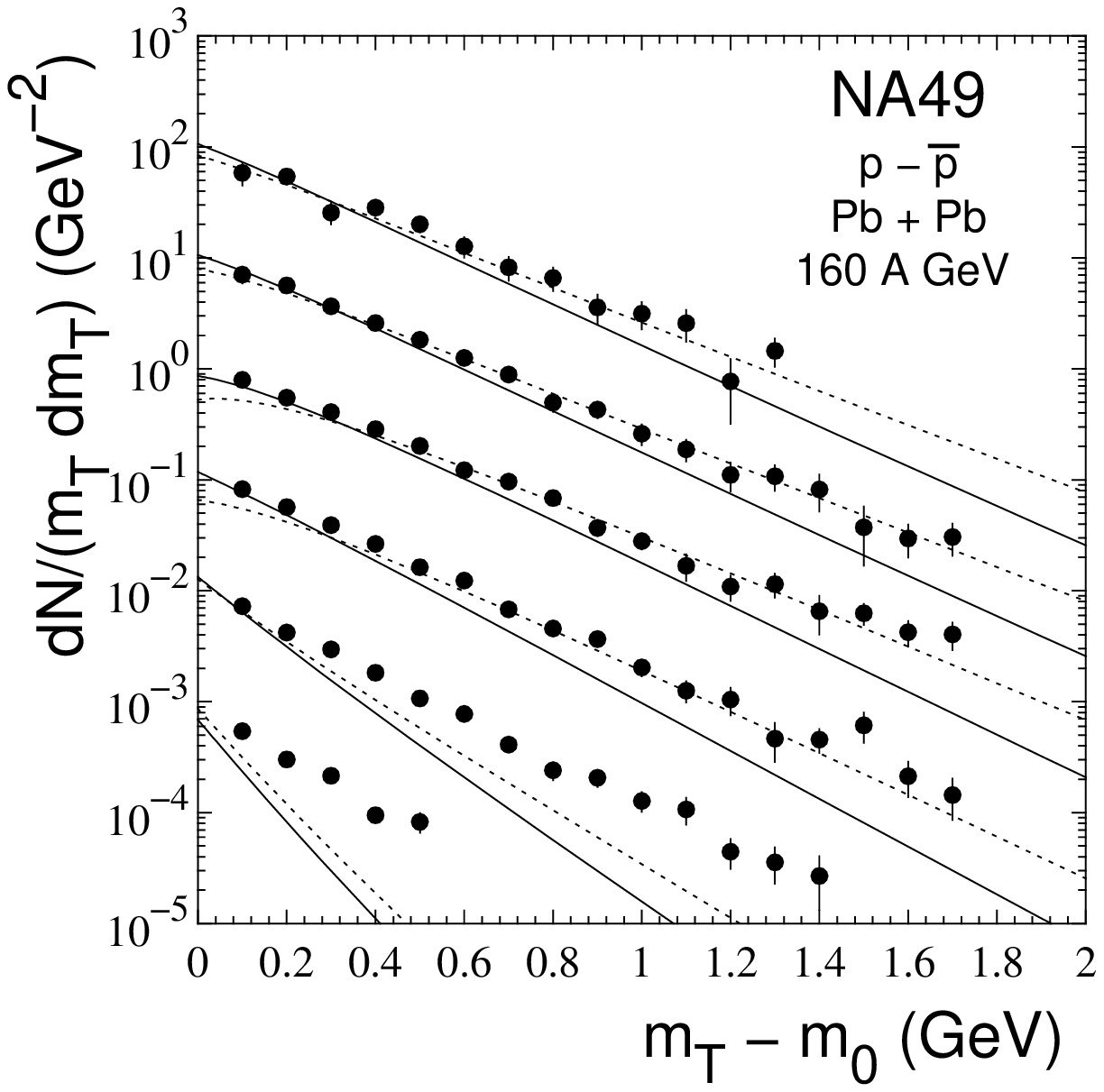}
        \hfill
  \end{minipage}\\

  \vspace*{-1.5truecm}
  \begin{minipage}[t]{\sizen}
        \epsfxsize \sizen \epsfbox{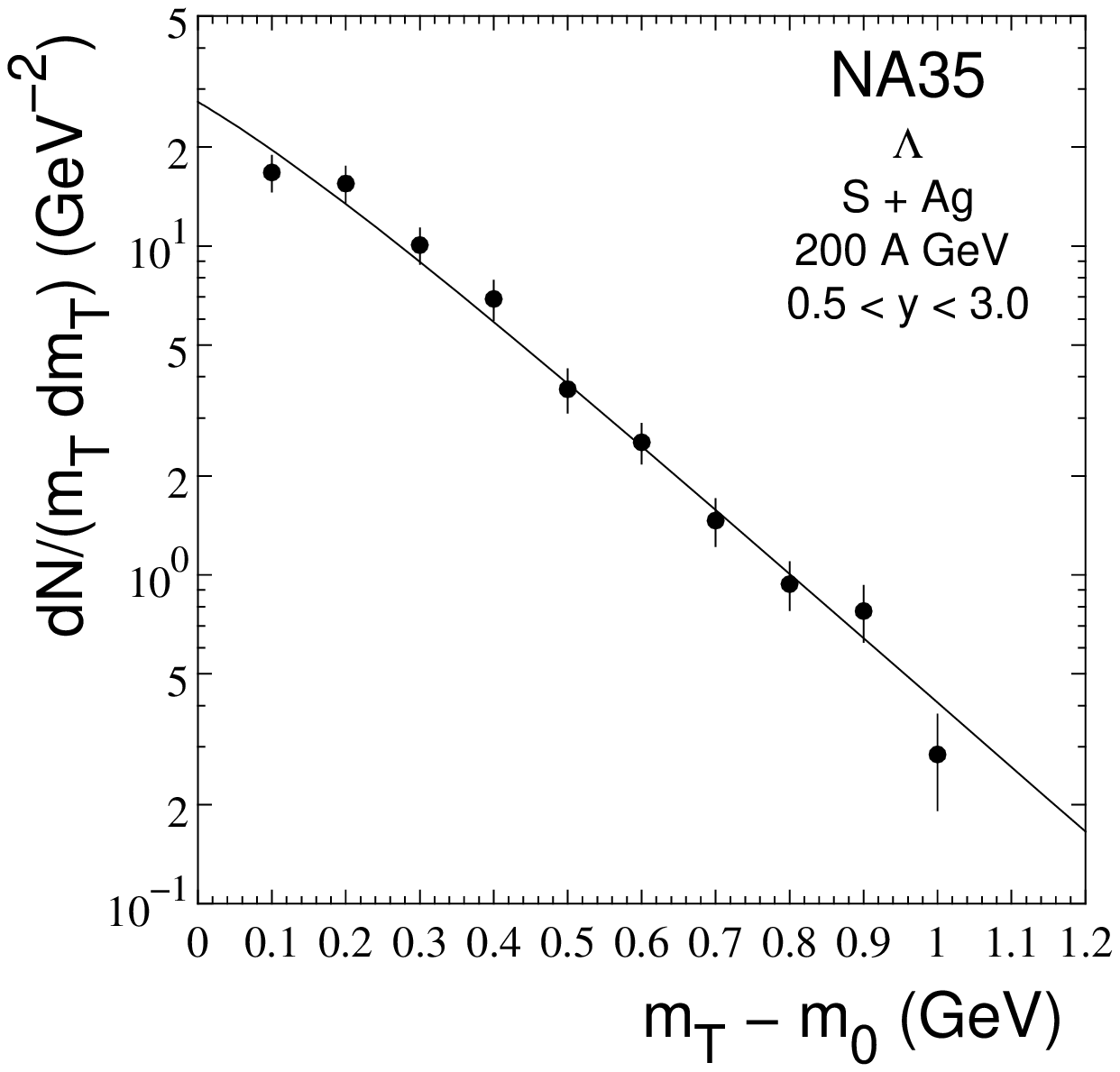}
        \hfill
  \end{minipage}
  \hfill
  \begin{minipage}[t]{\sizen}
        \epsfxsize \sizen \epsfbox{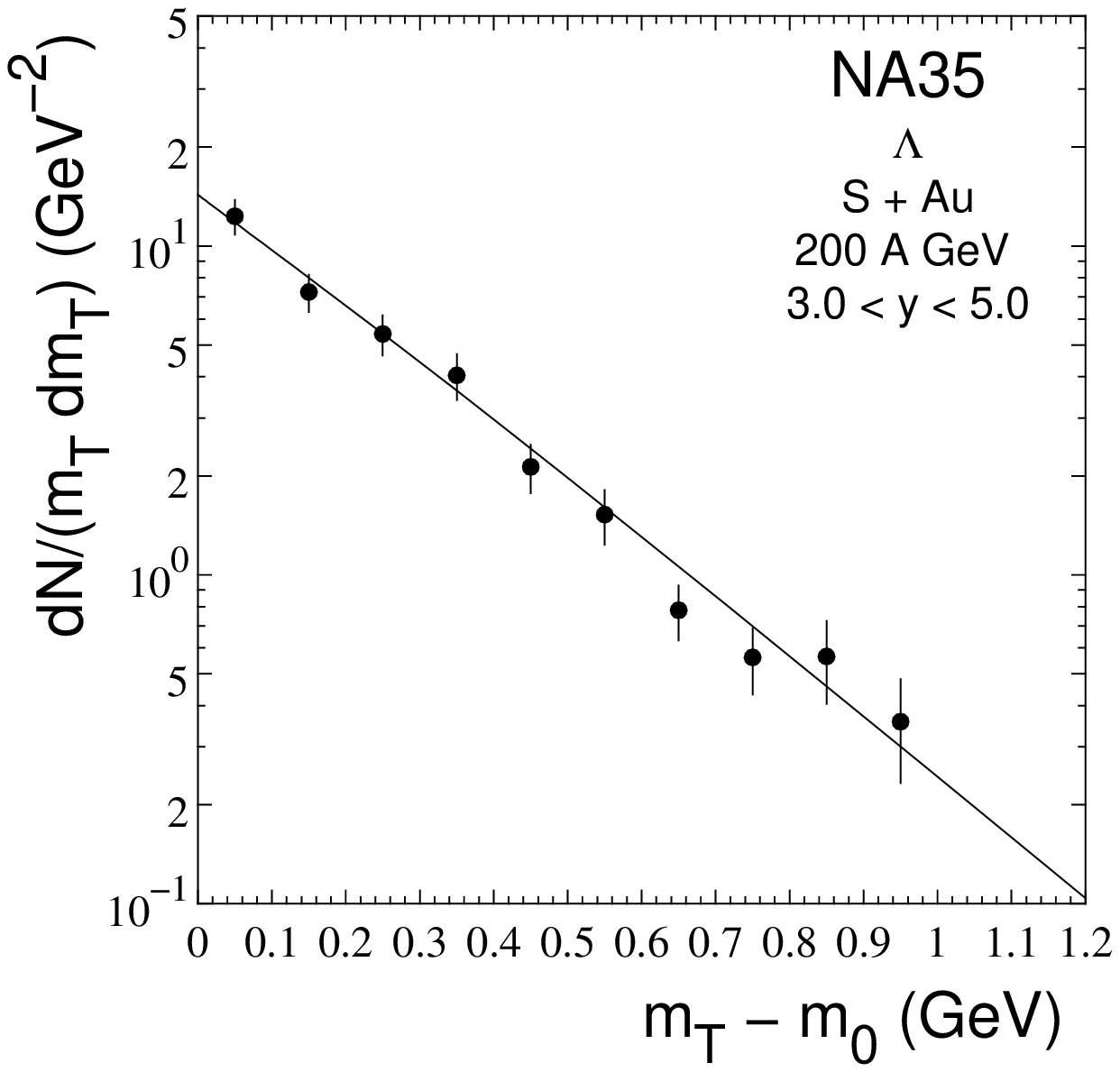}
        \hfill
  \end{minipage}

  {\bf Figure \ref{npropt}:}
\end{center}

\newpage

\begin{center}
   \begin{minipage}[t]{\size}
         \epsfxsize \size \epsfbox{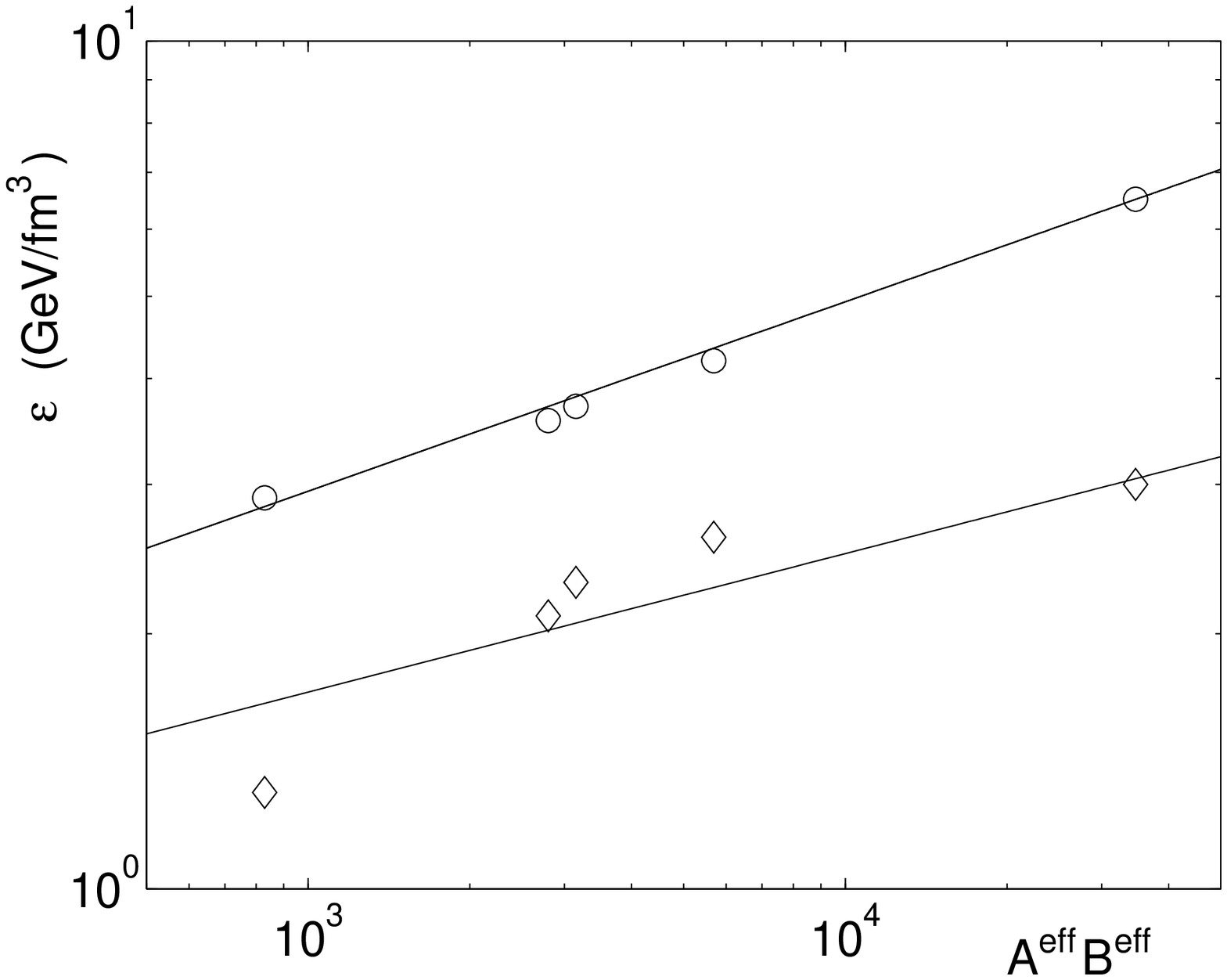}
         \hfill\\
   \end{minipage}\\

{\bf Figure \ref{sl}:}
\end{center}

   \refstepcounter{table}\label{tab1}
   \refstepcounter{figure}\label{Pblandau}
   \refstepcounter{figure}\label{inidens}
   \refstepcounter{figure}\label{inipro}
   \refstepcounter{figure}\label{freeze}
   \refstepcounter{figure}\label{negrap}
   \refstepcounter{figure}\label{nprorap}
   \refstepcounter{figure}\label{negpt}
   \refstepcounter{figure}\label{npropt}
   \refstepcounter{figure}\label{sl}

\end{document}